\documentclass{article}
 \usepackage[preprint]{log_2022}
\usepackage{booktabs}            % professional-quality tables
\usepackage{multirow}            % tabular cells spanning multiple rows
\usepackage{amsfonts}            % blackboard math symbols
\usepackage{graphicx}            % figures
\usepackage{duckuments}          % sample images

\usepackage{subcaption}
\usepackage{listings}
\usepackage{float}
\usepackage{enumitem}
\usepackage{xspace}
\usepackage{multirow}
\usepackage{array}
\usepackage{tabularx}
\usepackage{bm}
\usepackage{comment}
\usepackage{booktabs, makecell}
\usepackage{xcolor,colortbl}
\usepackage[english]{babel}
\usepackage{pifont} 
\usepackage{soul}
\usepackage{xfrac}
\usepackage{wrapfig}
\usepackage{mdwlist}
\usepackage{arydshln}
\usepackage{adjustbox}
\usepackage{graphicx,wrapfig,lipsum}
\usepackage{cellspace}
\usepackage{siunitx}
\usepackage{comment}

\newtheorem{theorem}{Theorem}

\usepackage{tikz}
\tikzstyle{xaxis-fill}=[rectangle,fill=black,text=white,minimum width=53em, minimum height=.65em, label={center:\scriptsize \bf \textcolor{white}{\textsf{1-Hop Sensitive Attribute Homophily}}}]
\usepackage{titlesec}
\titlespacing*{\subsubsection}{0pt}{.3\baselineskip}{.3\baselineskip}
\titlespacing*{\subsection}{0pt}{.4\baselineskip}{.3\baselineskip}
\titlespacing*{\section}{0pt}{.6\baselineskip}{.3\baselineskip}
\usepackage[linesnumbered,ruled,vlined]{algorithm2e}

\newcommand{\rebuttal}[1]{{\textcolor{black}{#1}}}

\usepackage[numbers,compress,sort]{natbib}
\title{On Performance Discrepancies Across Local Homophily Levels in Graph Neural Networks}

\author[D. Loveland et al.]{%
Donald Loveland\\
\institute{University of Michigan, Ann Arbor}\\
\email{dlovelan@umich.edu}\And
Jiong Zhu\\
\institute{University of Michigan, Ann Arbor}\\
\email{jiongzhu@umich.edu}\And
Mark Heimann\\
\institute{Lawrence Livermore National Lab}\\
\email{heimann2@llnl.gov}\And
Benjamin Fish\\
\institute{University of Michigan, Ann Arbor}\\
\email{benfish@umich.edu}\And
Michael T. Schaub\\
\institute{RWTH Aachen University}\\
\email{schaub@cs.rwth-aachen.de}\And
Danai Koutra\\
\institute{University of Michigan, Ann Arbor}\\
\email{dkoutra@umich.edu}
}

\begin{document}

\maketitle
\begin{abstract}
Graph Neural Network (GNN) research has highlighted a relationship between high homophily (i.e., the tendency of nodes of the same class to connect) and strong predictive performance in node classification. However, recent work has found the relationship to be more nuanced, demonstrating that simple GNNs can learn in certain heterophilous settings. To resolve these conflicting findings and align closer to real-world datasets, we go beyond the assumption of a global graph homophily level and study the performance of GNNs when the local homophily level of a node deviates from the global homophily level. Through theoretical and empirical analysis, we systematically demonstrate how shifts in local homophily can introduce performance degradation, leading to performance discrepancies across local homophily levels. 
We ground the practical implications of this work through granular analysis on five real-world datasets with varying global homophily levels, demonstrating that (a) GNNs can fail to generalize to test nodes that deviate from the global homophily of a graph, and (b) high local homophily does not necessarily confer high performance for a node. We further show that GNNs designed for globally heterophilous graphs can alleviate performance discrepancy by improving performance across local homophily levels, offering a new perspective on how these GNNs achieve stronger global performance. 
\end{abstract}

\section{Introduction}

Deep learning with Graph Neural Networks (GNNs) has become common in many learning tasks over collaboration networks \cite{cora_collab}, social networks \cite{Fan2019-GraphRec}, financial networks \cite{finance}, and more \cite{ciotti_homophily, wu2020connecting, vashishth2020compositionbased}. 
However, given the relative infancy of GNNs, retrospectives on GNN performance are limited. Understanding the conditions that will cause a GNN's performance to degrade promotes a proactive approach to GNN development, rectifying any issues that may arise once deployed to the public. One previously studied degradation mechanism of GNN performance is the presence of heterophilous connections \cite{zhu2020beyond, yan2021two}. 
Heterophily, or the tendency for nodes of different classes to connect, has been found in a variety of graph applications where sensitive factors may influence the connective patterns, necessitating its study \cite{hofstra2017sources, mcpherson2001birds, brea2018inference, karimi2018homophily}. 
However, more recent retrospectives have argued that performance does not necessarily degrade with heterophily, and in fact simple GNN architectures, such as GCN, can perform well in certain settings~\cite{ma2021homophily, luan2021heterophily, heterophily_ieeebul}. 
These seemingly conflicting results indicate a gap in understanding, demanding further research on the influence of heterophily on GNNs. 
%Furthermore, given that many sensitive factors can influence heterophily, one example being demographic information in social networks, it is important to deeply understand how GNNs make use of homophily and heterophily during predictions ~\cite{mcpherson2001birds, brea2018inference, karimi2018homophily}. 

To better understand the factors which govern GNN performance, we begin by exploring the assumptions made in previous works. Surprisingly, many assume that constituent nodes of a graph possess local homophily levels similar to the global homophily of the entire graph, leading to a disregard for the impact of a node's local homophily level on performance \cite{zhu2020beyond, ma2021homophily, luan2021heterophily}. 
For the few works that consider local homophily, it is often assumed that homophilous nodes should perform better, creating unclear conclusions due to biased interpretation of the localized results  \cite{du2022gbk, cavallo20222}. Moreover, local homophily has yet to be the focal point of previous works, leading to limited discussion on why certain patterns emerge. 
Practically, assuming a constant local homophily makes it difficult to determine if new models are improving performance across all nodes, or simply increasing performance for certain node subsets. Furthermore, by myopically assuming higher homophily is indicative of higher performance, analysis on discrepancies that arise from model choice, global homophily, and local homophily is limited, slowing the development of new GNNs that could address these concerns.
%That is, the source of disparity comes as a byproduct of the graph's generation process and GNN architecture, rather than an attribute of the node, and thus impacts any graph learning task.

%As such, implications also arise for fairness in human-centric applications as failing to characterize sources of disparity as local homophily varies can introduce systematic poor performance for individuals due to properties largely out of their control (e.g., graph generation and model assumptions).   

\vspace{0.1cm}
\noindent \textbf{This work.} 
%In this work, we investigate graphs where nodes have local homophily ratios which differ from the global homophily ratio, breaking the first assumption and aligning closer to real-world settings.
We investigate how shifts in local homophily can impact GNNs, extending beyond current assumptions and aligning closer to real-world settings.
Our analysis considers a GNN trained on nodes biased towards a graph's global homophily, and then applied to test nodes of varying local homophily levels. %Results are conveyed through groups of nodes that share similar local homophily ratios. 
We \textit{theoretically analyze} the scenario by obtaining a closed-form solution for a GNN's weight matrix and demonstrate, through perturbation analysis, that a GNN's performance can degrade when a node's local homophily level is shifted relative to the global homophily of the  graph. \rebuttal{We also theoretically demonstrate that when adopting aggregation mechanisms tailored for heterophilous graphs, predictions are less likely to degrade.} %Given the theoretical analysis is performed under ideal conditions, we also 
We show that these findings generalize to a variety of settings through a broad empirical analysis facilitated by our proposed \textit{synthetic graph generator} that enables control over the local homophily levels of nodes.  %within a generated graph. 
We also show the practical repercussions of our theoretical and empirical analyses on a representative set of five \textit{real-world} datasets with varying global homophily levels.
Across our synthetic and real-world datasets, we additionally study nine different GNN architectures, demonstrating that those tailored to handle heterophily often maintain more uniform performance, minimizing discrepancies. 
%Across our analysis, we find that the global homophily level can influence a GNN's ability to correctly classify a node, irrespective of local homophily, demonstrating that high local homophily confers higher performance only in certain settings.
%when there is a shift in local homophily. This phenomenon 
%presenting a challenge 
%for certain GNN models to 
%for nodes with underrepresented homophily levels to attain correct predictions. 
Together, our theoretical and empirical analysis describes a new failure point of GNNs -- an expected distribution of labels over a node's neighbors, stemming from an over-reliance on the global homophily of a graph -- presenting a challenge for nodes with underrepresented homophily levels to be correctly predicted. While previous works have noted fairness issues in GNNs based on sensitive attributes, e.g. race or gender, determined exogenously \cite{dong2022fairness, zhang2022fairness}, our results point to a novel inequality rooted in a network's structure that could lead to unfairness in human-centric settings. Our main contributions are: 
%in this work are summarized below: 

%points to how individuals dissimilar from the average population can be systematically treated different.

\vspace{-.1cm}
\setlist[itemize]{leftmargin=*}
\begin{itemize}%[noitemsep,topsep=1pt, wide] %,wide, labelwidth=!, labelindent=5pt]
    \item \textbf{Theoretical Analysis:} We show how a GNN's predictions change under a shift in local homophily level, providing intuition on how GNN performance can degrade for nodes with local homophily levels which differ from the global homophily level. \rebuttal{We analyze this premise for specific homophilous and heterophilous designs, showing that heterophilous GNNs will generally be less susceptible to performance degradation as a node's local homophily level varies.}
    %\item \textbf{Synthetic Generation Method: } \donald{We build upon the commonly used preferential attachment model, allowing for more granular control over the distribution of local homophily ratios in a graph. This capability allows for empirical verification of the theoretical insights in less restrictive settings.} 
    \item \textbf{Synthetic Experiments and Model Comparison:} We perform empirical analysis by modifying the preferential attachment model to allow for more granular control over the distribution of local homophily levels. This capability facilitates empirical verification of our theory under more general graph structures and GNN architectures. Additionally, we perform the first node-level analysis that directly compares GNNs that assume homophily and GNNs that are adjusted for heterophily, demonstrating different levels of performance discrepancy across GNN designs.

    %the first comparison of GNNs that rely on homophily and GNNs that are adjusted for heterophily in the context of shifts in local homophily, demonstrating different rates of degradation as local homophily levels deviate from the global homophily level. The analysis is facilitated through an extended preferential attachment model that allows for more granular control over the distribution of local homophily levels. This capability allows for empirical verification of the theoretical insights in less restrictive settings.
    \item \textbf{Real-world Experiments:} We provide the first granular analysis of GNN performance as local homophily levels are varied across a set of five real-world datasets. We find that our theoretical performance degradation trends hold more generally, confirming GNNs designed for heterophily can aid in minimizing performance discrepancy across nodes with varying local homophily patterns. 
    
\end{itemize}
\vspace{-.2cm}

\section{Related Work}

In this section, we begin by discussing GNN architectures designed to improve learning under heterophily. We then detail previous approaches towards local property and discrepancy analysis.
%, as well as discrepancy analysis. %, connecting both to our study of local homophily. 

\noindent \textbf{Learning GNNs in Diverse Neighborhoods.} GNNs adopt an aggregation function to combine the ego-node's (the node being updated) features and the neighboring nodes' features. Depending on the neighborhood of a node, a particular aggregation mechanism may be insufficient to learn representations \cite{yan2021two}. For example, GCN~\cite{kipf2016semi}, GAT~\cite{velickovic2018graph}, and SGC \cite{wu2019sgc} were built to learn over homophilous neighborhoods through their weighted average of the ego-node and neighboring nodes' features. To remedy this issue, models such as GraphSAGE~\cite{hamilton2017sage}, GPR-GNN \cite{chien2021adaptive}, FA-GCN \cite{bo2021fagcn}, GCNII \cite{chen2020gcnii}, \rebuttal{and LINKX \cite{lim2021large}} separate the ego and neighbor embeddings, either through a residual connection or concatenation. \rebuttal{We note that LINKX is MLP-based, directly embedding the adjacency matrix through an MLP rather than traditional message passing}. GPR-GNN and FA-GCN additionally follow a predict-before-propagate paradigm to help alleviate the harm that can come from mixing representation learning and aggregation \cite{Cui_2020, gasteiger2022predict}, while GCNII utilizes identity mapping to mitigate oversmoothing, a known problem for homophily \cite{yan2021two}. H2GCN adopts further decoupling across higher order neighborhoods, aggregating each $k$-hop neighborhood separately \cite{zhu2020beyond}. 
%GBK-GNN differentiates from other GNNs by introducing an architecture-agnostic learning mechanism that allows GNNs to fuse two different sets of weights, each intended to capture homophilous and heterophilous components of the data respectively~\cite{du2022gbk}. 
\textit{In previous works, there is limited analysis demonstrating the impact of GNN architectures on the performance of nodes with varying local homophily. In this work, we provide the first granular analysis of these models, showing a new perspective on how they perform and their (in)ability to mitigate performance discrepancies.}

\noindent \textbf{Local Property Analysis.}
Studies on GNN performance relative to an input graph's structural properties have gained traction; however, the adjustment to considering a per-node local perspective is still under-explored. For instance, many studies have argued the conditions in which a node is able to benefit from message passing with respect to homophily, but only consider a constant local homophily level for all nodes \cite{ma2021homophily, luan2021heterophily, yan2021two}. 
%In addition, both works only present aggregated real-world empirical results, obscuring what happens across the range of local homophily ratios. 
Du et al. offers the first local analysis, however the results are contradictory across datasets and are only performed for a single model \cite{du2022gbk}. More recent work has developed other homophily-inspired metrics to contextualize local performance, however the proposed metric can still fail to explain performance depending on the dataset \cite{cavallo20222}. Both works assume that higher local homophily should always improve performance, ultimately guiding their development of new architectures and metrics that reinforce this assumption. However, the conflicting results across datasets seen in both works indicate that this assumption may oversimplify the behavior of a GNN and fail to consider other drivers for performance degradation. Closely related to our work, Ma et al. analyzes the disparate treatment of individual nodes defined by their shortest path distance to the training dataset, showing a degradation in performance as distance increases \cite{ma2021subgroup}. We build upon the idea of structural property subgroup analysis, but instead consider variations in local homophily rather than distance to the training set, creating a shift in how performance is analyzed in the context of homophily. \textit{Thus, we analyze the performance of GNNs, breaking the assumption that the local homophily levels are constant, and demonstrate how node predictions systematically degrade as the local homophily levels deviate from the global homophily of the training graph.}

\section{Preliminaries}

In this section, we provide key notations and definitions, the notation is summarized in App. \ref{app:notation}.

\subsection{Graphs}

Let $G = (V, E, \mathbf{X}, \mathbf{Y})$ denote a simple graph with node set $V$ and edge set $E$, where $\mathbf{X} \in \mathbb{R}^{|V| \times f}$ represents the node feature matrix with $f$ features per node and $\mathbf{Y} \in \{0, 1\}^{|V| \times c}$ represents the one-hot encoded node label matrix with $c$ classes. A specific node $i \in G$ has feature vector $\mathbf{x_{i}}$, class $y_{i} \in \{1, ..., c\}$, and one-hot encoded class label vector $\mathbf{y_{i}}$. The edge set can also be represented as an adjacency matrix, $\mathbf{A} \in \{0, 1\}^{|V| \times |V|}$, where a value of $1$ at index $(i, j)$ denotes an edge between nodes $i$ and $j$ in $G$, otherwise the index is set to $0$. We use both $E$ and $\mathbf{A}$ throughout the paper, opting for $E$ when explicitly discussing the edges of $G$ and $\mathbf{A}$ when describing matrix computations on the edge set. A \textit{k-hop neighborhood} of node $i \in V$, $N_{k}(i)$,  denotes the subgraph induced by the nodes that are reachable within $k$-steps of $i$. 
%\reminder{move to 3.3? ->} The empirical class compatibility matrix of a graph, $[\mathbf{H}_{L}]$ describes the probability of two nodes with certain labels connecting, where the $(u, v)$-th entry is the fraction of edges between a node in class $u$ and a node in class $v$: $[\mathbf{H}_{L}]_{u, v} = \dfrac{|\{(i, j) : (i, j) \in E \land y_{i} = u \land y_{j} = v\}|}{|\{(i, j) : (i, j) \in E \land y_{i} = u \}|} $.

%It is formally defined as $[\mathbf{H}_{L}]_{u, v} = P\left((u, v) \in E | \mathrm{L}_u = l_u, \mathrm{L}_v = l_v\right) $ for two nodes $u$ and $v$, and can be empirically estimated. 

\subsection{Node Classification with GNNs}
We focus on node classification through a GNN, where the goal is to learn a mapping between $\mathbf{X}$ and $\mathbf{Y}$. This mapping is estimated through a subset of $V$, referred to as the set of training nodes $n_{train}$. For a $k$-layer GNN, learning is facilitated through message passing over $k$-hop neighborhoods of a graph. The steps, at a high level, include (1) embedding $\mathbf{X}$ through a non-linear transformation parameterized by a weight matrix $\mathbf{W}$ and (2) aggregating the embedded features across neighborhoods of each node.
Message passing over all nodes in the graph can be computed through matrix multiplication, where the most basic formulation updates node representations through  $\mathbf{R}_{l} = \sigma (\mathbf{(A + I)R}_{l-1}\mathbf{W}_{l})$ 
\begin{comment}
\ben{can't be right: $X is |V|\times f$, and $A$ is $|V|\times|V|$, so $W_i$ would be $f\times |V|$, but this should be the weights for a single node, no?  line 359 works out better in terms of dimensions} 
\end{comment}
for a layer $l \in \{1, 2, ..., k\}$ of the GNN, where $\mathbf{R}_{0} = \mathbf{X}$ and $\sigma$ is an activation function. The update is applied $k$ times, resulting in final representations for each node that can be used for classification.

\subsection{Homophily and Heterophily}

In this work, we focus on edge homophily and present the following definitions to describe our homophily-based analysis. We begin with the global homophily ratio of a graph, $h$.
\\[0.1cm]
\noindent \textbf{Definition 1 - Global Homophily Ratio.} 
\textit{The global homophily ratio $h$ over a graph's edge set $E$ is the fraction of edges in $E$ that connect two nodes, $u$ and $v$, with the same label, $y_{u}$ and $y_{v}$: }
\begin{equation}
h = \frac{|\{(u,v) : (u, v) \in E \land y_u = y_v\}|}{|E|}.
\label{eq:global_homophily_ratio}
% \vspace{-0.1cm}
\end{equation}
The global homophily ratio is used to describe the overall homophily level in graphs; $h = 0$ indicates a fully heterophilous graph and $h = 1$ indicates a fully homophilous graph \cite{mcpherson2001birds}. \rebuttal{Other global homophily metrics are discussed in Appendix \ref{app:other_metrics}.} Additionally, the empirical class compatibility matrix of a graph, $[\mathbf{H}_{L}]$ describes the probability of two nodes with certain labels connecting, where the $(u, v)$-th entry is the fraction of edges between a node in class $u$ and a node in class $v$: $[\mathbf{H}_{L}]_{u, v} = \dfrac{|\{(i, j) : (i, j) \in E \land y_{i} = u \land y_{j} = v\}|}{|\{(i, j) : (i, j) \in E \land y_{i} = u \}|} $. 
However, both the global homophily ratio and compatibility matrix oversimplify the mixing patterns in a graph when there are varying neighborhood compositions.
To perform more granular analysis on a per-node basis, we also define the local homophily ratio of a node $t$, $h_{t}$.

\vspace{0.05cm}
\noindent \textbf{Definition 2 - Local Homophily Ratio.} 
\textit{The local homophily ratio of a node $t$, $h_{t}$, is the fraction of edges in the neighborhood of $t$ that connect $t$ to a neighbor $u$ with the same class: }
\begin{equation}
h_{t} = \frac{|\{(u,t) : (u, t) \in N_{1}(t) \land y_u = y_{t}\}|}{|N_{1}(t)|}.
\label{eq:local_homophily_ratio}
% \vspace{-0.1cm}
\end{equation}
Given GNNs are often shallow and only depend on a small $k$-hop neighborhood for a single node prediction, it is natural to analyze GNNs through the local, rather than global, homophily ratio. Moreover, many real-world graphs display a wide range of local homophily ratios across the constituent nodes, as seen in App. \ref{app:real_distr}, necessitating local analysis.
%To facilitate our study, we group test nodes together based on similar local homophily ratios and measure their collective performance.

\section{Relationship between a Node's Local Homophily Level and Performance}
\label{theory_section}

%While some previous works have explored discrepancy in performance across local homophily ranges, many results have been purely empirical. Moreover, many of the empirical analyses have been been largely inconclusive or under-explored since they are typically supplementary rather than the key focus of the work. Thus, 
In this section, we aim to characterize the impact of local homophily on the accuracy of node-level predictions by considering shifts in local homophily levels, at test time, relative to the global homophily level the GNN was trained on. We begin by revealing the drivers for performance discrepancies through theoretical analysis and discuss their implications on node-level performance. Leveraging these insights, we relax our assumptions in Section \ref{synth_section} and show that our theory holds in more general settings via extensive empirical analysis on synthetic data. Additionally, we consider even more general real-world graphs (without any constraints) in Section~\ref{real_section}. 

%Then, we relax our assumptions and show that our theory holds in more settings via extensive empirical analysis on synthetic data in Section \ref{synth_section}, as well as through more general real-world data in Section \ref{real_section}.

\textbf{Setup.} %We set up our theory similar to Zhu et al. \cite{zhu2021relationship}. 
Following previous theoretical GNN work~\cite{zhu2020beyond, du2022gbk, wang_fairgnn, luan2021heterophily} and popular models such as SGC and LightGCN~\cite{wu2019sgc, he2020lightgcn}, we study 
% a similar structure to the GNNs studied in previous theoretical work~\cite{zhu2020beyond, du2022gbk, wang_fairgnn, luan2021heterophily}, and is similar to both SGC and LightGCN~\cite{wu2019sgc, he2020lightgcn}.
% We assume 
\rebuttal{two different GNNs with different aggregation mechanisms. Our homophilous GNN}, $F$, is formulated as $\mathbf{(A + I)XW}$, where $\mathbf{A + I}$ is $G$'s adjacency matrix with self-loops, %and $\mathbf{W}$ is $F$'s weight matrix, 
\rebuttal{directly mixing the features of the ego-node and neighbor nodes \cite{kipf2016semi, wu2019sgc, he2020lightgcn}. Our heterophilous GNN, $F'$ is formulated as $\mathbf{(X \mathbin\Vert AX)W}$, where $\mathbf{(X \mathbin\Vert AX)}$ is the concatenation of the ego-node features and aggregated neighboring features \cite{hamilton2017sage, zhu2020beyond, lim2021large}. 
%$F$ mixes the features of the ego-node and neighbor nodes as seen in popular homophilous models \cite{kipf2016semi, wu2019sgc, he2020lightgcn}, 
%$F'$'s aggregation technique has been shown to improve learning under heterophily over the weighted aggregation with self-loops \cite{hamilton2017sage, zhu2020beyond, lim2021large}. 
Both designs have yet to be analyzed through a localized perspective to characterize discrepancy across predictions.} 

Similar to the setup in~\cite{zhu2021relationship}, we consider a graph $G$ with a subset of training nodes $n_{train}$, each of which %. Each node $i \in n_{train}$ 
has an associated node feature vector $\mathbf{x_{i}}$, one-hot encoded class label vector $\mathbf{y_{i}}$, 1-hop homophily ratio $h$, and degree $d$. For brevity, we focus on binary classification (though we consider multi-class settings in our experiments) and represent $\mathbf{y_{i}}$= $onehot(y_i) = \begin{bmatrix} 1 & 0 \end{bmatrix}$ when node i's class is $y_i = 0$ and $\begin{bmatrix} 0 & 1 \end{bmatrix}$ when $y_i = 1$. 
We consider node feature vectors from a uniform distribution and  
% We treat the node feature vectors as uniform vectors 
biased towards a particular class:  % by $p \in [0, 0.5]$, 
 when $y_{i} = 0, \mathbf{x_{i}} = \begin{bmatrix} (0.5+p) & (0.5-p) \end{bmatrix}$ and when $y_{i} = 1, \mathbf{x_{i}} = \begin{bmatrix} (0.5-p) & (0.5+p) \end{bmatrix}$, where parameter $p \in [0, 0.5]$ controls the `agreement' between the node features and its class label, \rebuttal{i.e. as $p$ approach $0.5$ the features become more similar to the class labels}.
% Following previous theoretical GNN work~\cite{zhu2020beyond, du2022gbk, wang_fairgnn, luan2021heterophily} and popular models, SGC and LightGCN~\cite{wu2019sgc, he2020lightgcn}, we assume 
% a similar structure to the GNNs studied in previous theoretical work~\cite{zhu2020beyond, du2022gbk, wang_fairgnn, luan2021heterophily}, and is similar to both SGC and LightGCN~\cite{wu2019sgc, he2020lightgcn}.
% We assume 
% a simple GNN, $F$, formulated as $\mathbf{(A + I)XW}$, where $\mathbf{A + I}$ is $G$'s adjacency matrix with self-loops and $\mathbf{W}$ is $F$'s weight matrix, is trained on $n_{train}$. 
%We note that $F$ follows a similar structure to the GNNs studied in previous theoretical work \cite{zhu2020beyond, du2022gbk, wang_fairgnn, luan2021heterophily}, and is similar to both SGC and LightGCN \cite{wu2019sgc, he2020lightgcn}. 
The final prediction for node $i$ is $\mathbf{argmax} \, \mathbf{{z}_{i}}$ where $\mathbf{{z}_{i}}$ is the output logit vector of the GNNs. 
We begin by solving for both GNN's optimal weight matrix $\textbf{W}$, and then apply $F$ \rebuttal{and $F'$} to a test node $t$. We specify the local homophily ratio for $t$ as $h + \alpha_{t} = h_{t}$, where $\alpha_{t} \in [-h, 1-h]$ is $t$'s shift in local homophily level compared to the global homophily level. Under this setup, we analyze how  $t$'s prediction is impacted as its local homophily ratio $h_{t}$ shifts relative to $h$, the global homophily ratio used to train $F$. In Theorem 1, without loss of generality, we consider the impacts of $h_{t}$ when $t$'s class label is $y_{t} = 0$ \rebuttal{for GNN $F$. In Theorem 2, we consider the same setup for GNN $F'$. } 

\vspace{-.4cm}
\textbf{Analysis \& Implications.} Our first theorem provides a \textit{direct relationship between $t$'s performance \rebuttal{under a homophilous GNN} and its local homophily ratio} as it deviates from the global homophily.

\vspace{-0.1cm}
\begin{theorem}
Consider a test node $t$ with local homophily ratio $h + \alpha_{t}$, label $\mathbf{y_{t}} = \begin{bmatrix} 1 & 0 \end{bmatrix}$, and node features $\mathbf{x_{t}} = \begin{bmatrix} (0.5+p) & (0.5-p) \end{bmatrix}$. The class prediction from $F$ for node $t$ is a function of the global homophily level and the shift of the local homophily level, given by  $\mathbf{argmax}  \, \mathbf{z_{t}}$, where $\mathbf{z_{t}} = \mathbf{y_{t}} + b_{1}\begin{bmatrix} \alpha_{t} & -\alpha_{t} \end{bmatrix}$ and $b_{1} = d/(1+d(2h-1))$. 

%produced by a GNN with weight matrix $\textbf{W}$ and the aforementioned architecture, is given by $\mathbf{z_{t}} = \mathbf{y_{t, 0}} + b\begin{bmatrix} \alpha_{t} & -\alpha_{t} \end{bmatrix}$ where $b = \dfrac{d}{1+d(2h-1)}$.

%The logit vector $\mathbf{z}_{t}$ for a test node $t$ with local homophily ratio $h + \alpha_{t}$, node label $\mathbf{y_{t,0}} = \begin{bmatrix} 1 & 0 \end{bmatrix}$, and node feature vector $x_{t} = \dfrac{1}{2} \begin{bmatrix} (1+p) & (1-p) \end{bmatrix}$, produced by a GNN with weight matrix $\textbf{W}$ and the aforementioned architecture, is given by $\mathbf{z_{t}} = \mathbf{y_{t, 0}} + b\begin{bmatrix} \alpha_{t} & -\alpha_{t} \end{bmatrix}$ where $b = \dfrac{d}{1+d(2h-1)}$.
\end{theorem}

\textbf{Proof. } The proof can be found in App.~\ref{app:proof}. We provide additional analysis on a 2-layer variant of $F$, formulated as $\mathbf{(A + I)^{2}XW}$ in App. \ref{app:2_layer_proof}.

    % \textbf{\textit{Homophilous} GNNs are Susceptible to Performance Discrepancies.} Theorem 1 provides a \textit{direct relationship between $t$'s performance \rebuttal{under a homophilous GNN} and its local homophily ratio} as it deviates from the global homophily ratio.
    Intuitively, this theorem implies that \textbf{\textit{homophilous} GNNs are susceptible to performance discrepancies.} 
    Specifically, we can expect the \textit{performance to degrade} when a test node either becomes \textit{more homophilous relative to a heterophilous graph} or  \textit{more heterophilous relative to a homophilous graph}. To further understand the implications of $\alpha_{t}$, we analyze three settings that naturally arise for the global homophily: (1)~$0 \le h < 0.5$, (2)~$h = 0.5$, and (3)~$0.5 < h \le 1$. We note that while the node degrees can influence the conditions that cause $\alpha_{t}$ to impact $\mathbf{z_{t}}$ (through $b_{1}$), we show in App.~\ref{app:proof} that this mostly occurs for extremely low-degree nodes. Previous work corroborates our findings regarding the difficulty with low-degree nodes under heterophily~\cite{zhu2020beyond, yan2021two, bei2023_low_node_diff}; however, our analysis extends this observation to demonstrate a significantly more complex interplay between node degree, global homophily, and shift in local homophily. 

\noindent \textbf{Setting 1: Heterophilous $(\mathbf{0 \le \textit{h} < 0.5}$)}: In this scenario, when $d(2h - 1) < -1$, $b_{1} < 0$, leading to $\mathbf{z_{t}} =  \mathbf{y_{t}} + |b_{1}|\begin{bmatrix} -\alpha_{t} & \alpha_{t} \end{bmatrix}$, where  $b_{1}$'s sign has been distributed into the vector. Thus, 
\begin{equation} \label{eq:case1}
\mathbf{z_{t}} = 
    \begin{cases}
      \mathbf{y_{t}} + |b_{1}|\begin{bmatrix} |\alpha_{t}| & -|\alpha_{t}| \end{bmatrix},  & \text{if}\ h_{t} \le h \\
      \mathbf{y_{t}} + |b_{1}|\begin{bmatrix} -\alpha_{t} & \alpha_{t} \end{bmatrix},  & \text{if}\ h_{t} > h ,
    \end{cases}
\end{equation}
where the sign of $\alpha_{t}$ has been integrated into the vectors. We can then deduce that (globally) heterophilous graphs, when $b_{1} < 0$ is satisfied, \textbf{will cause $F$ to degrade in performance as the test node's local homophily increases}, denoted by the score increase of the wrong class in the second case of Equation {\eqref{eq:case1}}. Additionally, when local homophily decreases, the predictions will improve given the increase in score for the correct class of the first case in Equation \eqref{eq:case1},
\begin{comment}
\ben{something has to be wrong here.  when $h_t < h$, so $\alpha < 0$, and $b$ is negative, then decreasing local homophily (below $h$, which is presumably what you mean here) will decrease $1+b(-\alpha)$, i.e. move it away from $1$ which means a worse prediction, not a better prediction}
\end{comment} 
however as $h < 0.5$, $\alpha_{t}$ has a smaller possible range of values, minimizing the impact on $F$'s predictions.

\vspace{0.1cm}
\noindent \textbf{Setting 2: Mixed Homophily ($\mathbf{\textit{h} = 0.5}$)}: When the graph is not strongly homophilous nor strongly heterophilous (i.e., $h=0.5$), $b_{1} = d$, leading to:
    \begin{equation}
    \mathbf{z_{t}} = 
        \begin{cases}
          \mathbf{y_{t}} + d\begin{bmatrix} -|\alpha_{t}| & |\alpha_{t}| \end{bmatrix},  & \text{if}\ h_{t} \le 0.5  \\
          \mathbf{y_{t}} + d\begin{bmatrix} \alpha_{t} & -\alpha_{t} \end{bmatrix}, & \text{if}\ h_{t} > 0.5 .
        \end{cases}
    \end{equation}
In this case, we find that $F$ will have improved performance when the local homophily of a test node is increased. Conversely, decreased local homophily for a test node will decrease performance. This is the only case that agrees with previous work regarding high homophily as a direct indicator of performance. Notably, the prediction directly depends on $d$, potentially leading to performance variations that are dominated by degree, rather than local homophily.

\vspace{0.1cm}
\noindent \textbf{Setting 3: Homophilous ($\mathbf{0.5 < \textit{h} \le 1}$)}: In this scenario, $b_{1} > 0$, leading to:
\begin{equation}
    \mathbf{z_{t}} = 
        \begin{cases}
          \mathbf{y_{t}} + |b_{1}|\begin{bmatrix} -|\alpha_{t}| & |\alpha_{t}| \end{bmatrix},  & \text{if}\ h_{t} \le h \\
          \mathbf{y_{t}} + |b_{1}|\begin{bmatrix} \alpha_{t} & -\alpha_{t} \end{bmatrix},  & \text{if}\ h_{t} > h .
        \end{cases}
    \label{eq:case3}
    \end{equation}
We can then deduce that (globally) homophilous graphs \textbf{will cause $F$ to degrade in performance as a test node's local homophily decreases}, denoted by the score increase of the first case of Equation~\eqref{eq:case3}. When local homophily increases, the predictions will improve, however as $h > 0.5$, $\alpha_{t}$ has a smaller range of values, minimizing impact on the predictions. 

\rebuttal{Going beyond homophilous GNNs,  Theorem 2 provides a \textit{direct relationship between $t$'s performance under a heterophilous GNN and its local homophily ratio} as it deviates from the global homophily ratio. 
At a high level, it 
implies that  \textbf{\textit{heterophilous} GNNs help alleviate performance discrepancies.}}

\vspace{-.4cm}
\rebuttal{
\begin{theorem}
Following the same setup from Theorem 1, the class prediction from $F'$ for node $t$ is a function of the global homophily level and the shift of the local homophily level, given by  $\mathbf{argmax}  \, \mathbf{z_{t}}$, where $\mathbf{z_{t}} = \mathbf{y_{t}} + b'_{1}\begin{bmatrix} \alpha_{t} & -\alpha_{t} \end{bmatrix}$ and $b'_{1} =  d^2 (2h - 1)/(1 + d^2(2h - 1)^{2})$. 
\end{theorem}
}
\rebuttal{
\textbf{Proof. } The proof can be found in App.~\ref{app:hetero_proof}.
}

\rebuttal{
% \textbf{\textit{Heterophilous} GNNs Help Alleviate Performance Discrepancies.} 
% At a high level, Theorem 2 provides a \textit{direct relationship between $t$'s performance under a heterophilous GNN and its local homophily ratio} as it deviates from the global homophily ratio. 
Comparing to Theorem 1, we can \textit{still expect performance degradation}, although at a different rate due to $b'_1$. Our goal is to determine which of the two models produces the largest perturbations to the output logit vector, $\mathbf{z_{t}}$, introducing more significant discrepancies. In Figure \ref{diff_coeff} in App. \ref{app:hetero_proof}, we compare $b'_1$ and $b_{1}$, the coefficients on the vector $\begin{bmatrix} \alpha_{t} & -\alpha_{t} \end{bmatrix}$ for the heterophilous and homophilous GNNs, respectively. We find that the magnitude of $b'_{1}$ is often significantly smaller than the magnitude of $b_{1}$ when varying values of $d$ and $h$. Thus, \textit{we can expect $\begin{bmatrix} \alpha_{t} & -\alpha_{t} \end{bmatrix}$ to have less impact on $\mathbf{z_{t}}$ for heterophilous models}, creating less discrepancy as compared to homophilous models. The full analysis comparing $b_{1}$ and $b'_{1}$ can be found in App.~\ref{app:hetero_proof}. 
}

\vspace{0.1cm}

%\noindent \textbf{Remark. } Previous work has observed that low-degree nodes can complicate learning for graphs with heterophily~\cite{zhu2020beyond, yan2021two}; however, our analysis extends this observation to demonstrate a significantly more complex interplay between node degree (as discussed in App.~\ref{app:proof}), global homophily, and shift in local homophily. %Furthermore, while our theory specifies certain conditions, our subsequent empirical analysis demonstrates the generalization of our findings.

%%%%%%%%%%%%%%%%%%%%%%%%%%%%%%%%%%%%%%%%%%%

\section{Generalization of Theoretical Results via Synthetic Data Analysis}
\label{synth_generation_section}

To display how the theoretical relationship between local homophily and classification performance generalizes, we introduce a graph generator that enables control over the local homophily ratios across a graph and conduct an extensive empirical analysis to study the following research questions: \textbf{(Q1)}~What performance discrepancies arise across the range of local homophily values as the global homophily is varied? and \textbf{(Q2)}~Do GNNs built specifically for heterophily display different discrepancy patterns across local homophily ranges as compared to homophilous architectures?

\subsection{Synthetic Data Generation} 

Building on the preferential attachment model where a compatibility matrix governs edge likelihood~\cite{karimi2018homophily, zhu2020beyond}, we modify the generator to allow a node's homophily level to be either randomly assigned or defined by the compatibility matrix. \rebuttal{We explain the generation process below, and provide the explicit steps 
in Algorithm \ref{synth_gen}, found in App. \ref{app:synth_dataset_details}}. At a high level, the steps to add a node to a synthetic graph are: (1) Sample a class label, (2) Generate node features, and (3) Add connections based on assigned homophily level. The code is available in an anonymized repository\footnote{\url{https://anonymous.4open.science/r/HeterophilyDiscrepancyGNN-85FB}}. We provide related work for graph generation in App. \ref{app:synth_dataset_related}, and property analysis in App. \ref{app:synth_dataset_props}.

%Given the interest in studying how variations in the local homophily distribution can impact GNN learning, we require a method that allows us to systematically modify the distribution of local homophily ratios in a controlled manner.

\vspace{0.1cm}
\noindent \textbf{Class and Feature Generation.} For a node $i$, label $y_{i}$ is sampled from probability distribution $P(\{0, ..., c\})$ with $c$ possible classes. Features $\mathbf{x_{i}}$ are generated from a 2D Gaussian, where each dimension has a mean of $\epsilon y_{i}$ and standard deviation of 1, where $\epsilon \in [0, 1]$ introduces noise into the features. When $\epsilon = 1$, the label $y_{i}$ is explicitly encoded, when $\epsilon = 0$, $y_{i}$ is unrecoverable.

\vspace{0.1cm}
\noindent \textbf{Structure Generation.} 
% User-defined $h_{c}$ and $h_{s}$ define the probability of connecting to a node of similar property (homophilous) in each matrix, and thus the diagonal entries of each matrix is set to $h_{c}$ and $h_{s}$. 
Similar to previous heterophily  analyses, we define a class compatibility matrix $\mathbf{H}_{L}$ with diagonal elements $h$, denoting the probability of connecting nodes with similar classes (homophilous), and off-diagonals elements $(1-h)/c$, denoting the probability of connecting nodes with different classes (heterophilous) \cite{karimi2018homophily, zhu2020beyond}. During generation, a new node $u$ is attached to an existing node $v$ as $P\left((u, v) \in E\right) \propto [\mathbf{H}_{L}]_{y_{u}, y_{v}}$. To control the local homophily ratios, we introduce a uniformity parameter $\rho$ such that with probability $\rho$, a node $i$'s local homophily ratio $h_{i}$ is sampled at random from a uniform distribution, $U(0, 1)$. As $\rho$ increases, more local homophily ratios follow the random distribution, rather than the compatibility matrix. Since the preferential attachment model adds nodes sequentially, it is possible that the local homophily of nodes early in the generation process drift from their original values. \rebuttal{For example, if a node $i$ is initially generated with two homophilous connections and three heterophilous connections ($h_{i} = 0.4$), later nodes attached to $i$ can result in a final $h_{i} \neq 0.4$}. To correct this, we keep track of how far a node $i$ has drifted from its original $h_{i}$ through $drift_{i}$, and prioritize connections to high-drift nodes (i.e., a node $drift_{i} > \delta$, where $\delta$ is a hyperparameter that defines the drift threshold) that would return the node's local homophily ratio back to its original value. %\rebuttal{The values $drift_{i}$ to $drift_{n}$, where $n$ is the number of nodes in the graph, are stored in a vector $drift$.} % set during the node's creation. 

\setlength{\textfloatsep}{0.1cm}
\setlength{\floatsep}{0.1cm}

%\footnotetext{$G - sigdrift$ denotes nodes in $G$ not in $sigdrift$.}

\subsection{Synthetic Evaluation Setup}
\label{synth_section}
%Our synthetic empirical analysis aims to determine the following research questions: \textbf{(Q1)} What performance disparities arise across the range of local homophily values as the global homophily is varied? and \textbf{(Q2)} Do GNNs built specifically for heterophily display different performance patterns across varying local homophily ratio ranges as compared to simpler GNN architectures?
We begin by detailing the GNNs and  the synthetic graph generation process used in our experiments.
%used to probe these questions. 

\vspace{0.1cm}
\noindent \textbf{GNN Models.} For our experiments, we consider a diverse set of GNN models including SGC~\cite{wu2019sgc}, GCN~\cite{kipf2016semi}, GAT \cite{velickovic2018graph}, GraphSAGE \cite{bo2021fagcn}, GCNII \cite{chen2020gcnii}, H2GCN \cite{zhu2020beyond}, GPR-GNN~\cite{chien2021adaptive}, FA-GCN~\cite{bo2021fagcn}, \rebuttal{and LINKX~\cite{lim2021large}}. SGC, GCN and GAT act as homophilous baselines, while GCNII, H2GCN, GPR-GNN, FA-GCN, \rebuttal{and LINKX} adopt mechanisms to improve learning in heterophilous settings. While GraphSAGE was not built for heterophily, later works~\cite{zhu2020beyond} have noted its design can improve learning over heterophily;  we include it to represent a midpoint between the models. We also include results for a graph-agnostic MLP. Experiments across SGC \cite{wu2019sgc}, GPR-GNN \cite{chien2021adaptive}, and FA-GCN \cite{bo2021fagcn} can be found in Appendix \ref{app:synth_extra}. All models are hyperparameter tuned (parameters detailed in App \ref{app:hyp_param}) and the model which maximizes performance on the validation set is applied to the test set.

\begin{comment}
~\mh{maybe we have to explain that we characterize GraphSAGE as heterophily-inspired since it contains some of the designs proposed in ~\cite{zhu2020homophily}, since some reviewers may note that GraphSAGE wasn't originally proposed to tackle this problem}. 
\end{comment}

\begin{figure}[t!]
    \centering
  
    \includegraphics[ width=0.85\textwidth, keepaspectratio]{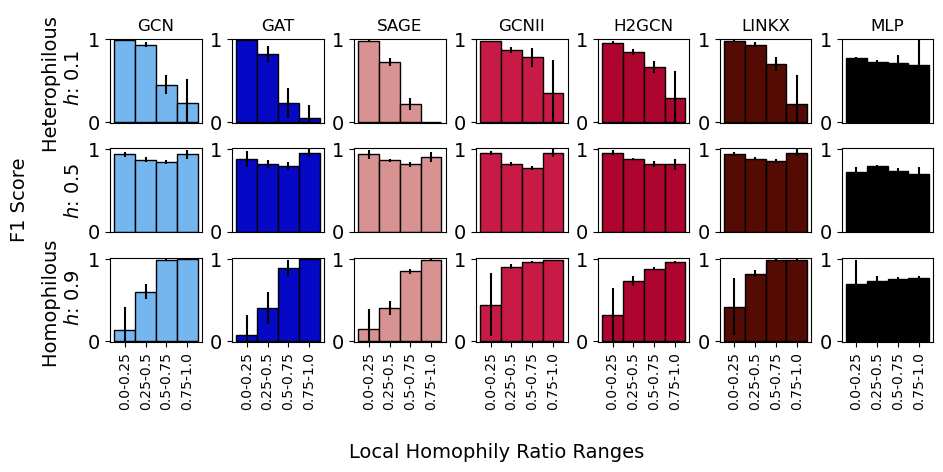}
    \vspace{-0.15cm}
    \caption{Performance for different GNN models (columns) on synthetic data generated with global homophily ratios $h \in \{0.1, 0.5, 0.9\}$ (rows) and uniformity $\rho = 0.5$, reported for different local homophily ranges. 
    %Each bar represents the F1 score for nodes with a local homophily ratio between the values on the x-axis. 
    % Error bars denote the standard deviation in performance. 
    % Blues indicate models with homophilous designs, reds indicates models with heterophilous designs. 
    As the local homophily deviates from the global homophily, homophilous models (blue bars) fail to generalize, resulting in performance discrepancy, while heterophilous models (red bars) retain \rebuttal{higher relative performance by up to 0.5 F1-score, alleviating discrepancy. Exact differences can be found in Figure \ref{fig:pairwise_diff_synth}}. MLP achieves constant performance.}
    \label{fig:local_vs_global_homophily}
    \vspace{-.1cm}
\end{figure}

\noindent \textbf{Data and Evaluation.} 
Using the proposed generator, we generate graphs by varying the global homophily ratio $h \in \{0.1, 0.3, 0.5, 0.7, 0.9\}$ with $\rho = 0.5$, $\epsilon = 0.5$, and $\delta = 5$, allowing us to study GNNs while jointly varying global and local homophily levels. An additional study for $\rho = 0.75$ is provided in Figure \ref{fig:synth_diff_uniform} of App. \ref{app:synth_extra}, demonstrating how discrepancy can be alleviated when the entire range of local homophily ratios has a sufficient number of nodes. %and explore the conditions determined by the theoretical analysis. 
%For feature generation, we set $\epsilon = 0.5$ to be sure the model cannot solely rely on the node features to make a prediction. 
For each combination of $h$ and $\rho$, we generate 10 graph with 5000 nodes and 100k edges (i.e. n = 5000, m = 20), and split the nodes into a 50-25-25\% split for train, validation, and test. \rebuttal{Additional discussion and experiments on varying the training set size are provided in Appendix \ref{app:synth_extra}}. To match our theoretical analysis, we focus on binary classification.
%Final structural details of the generated graphs are detailed in Appendix \ref{app:synth_dataset_props}.
For evaluation, we compare the performance across models and global homophily ratios as the local homophily ratio is varied. To compute localized performance, we split the test nodes into four groups based on their local homophily ratios and calculate an F1 score per group. \rebuttal{We choose F1 score as the node subsets can become imbalanced with respect to their class after splitting.} Per bin, we report the average and standard deviation for F1 over the 10 generated graphs. Results are presented in Figures \ref{fig:local_vs_global_homophily} of the main text, and Figures \ref{fig:local_vs_global_homophily_37}, \ref{fig:synth_extras} in App. \ref{app:synth_extra}

% \subsection{Synthetic Experiments}

\subsection{Synthetic Results}

\noindent \textbf{(Q1) Performance discrepancies across local and global homophily ratios.}
In Figure \ref{fig:local_vs_global_homophily}, we visualize the performance of each model by binning nodes of similar homophily for each of the global homophily ratios (analysis on $h \in \{0.3, 0.7\}$ and additional GNNs in App \ref{app:synth_extra}). We first find that the MLP, as expected, performs significantly worse with an F1 score of 0.75 as compared to the GNN's which achieve an F1 score of 0.9 and higher, highlighting the positive contribution of the graph structure on predictions. Next, we highlight the clear trend between performance and local homophily depending on the global homophily ratio: when $h < 0.5$ (lower global homophily), performance often degrades as local homophily increases, while when $h > 0.5$ (higher global homophily), performance often degrades as local homophily decreases. 
The results align well with our theoretical analysis, showing strong generalization of our findings to more complex graph structures and GNN architectures. Together, \textbf{our theoretical and empirical results indicate that assuming high homophily always indicates high performance may oversimplify the GNN's behavior, leading retrospective analyses astray when diagnosing performance degradation.} 
%Despite the trends when $h \ne 0.5$, the synthetic graphs with $h = 0.5$ demonstrate less pronounced performance variations. It is possible that GNNs in practice are less likely to rely on a global homophily ratio of $0.5$ given the majority of the local neighborhoods in this graph will have an even mix of homophilous and heterophilous connections. However, as the neighborhoods become either more heterophilic or homophilic, the model is able to once again leverage neighborhood information to improve performance, albeit ending at a lower F1 than nodes of similar local homophily in graphs of global homophily close to the local homophily value. 

\begin{comment}
\begin{figure}[t!]
    \centering
    \includegraphics[width=0.37\textwidth, keepaspectratio]{KDD2023_Figures/subset_synth_model_diff_cr.png}
    \vspace{-0.3cm}
    \caption{Pairwise performance differences for synthetic graphs ($h \in \{0.1, 0.9\}$) over GNN models identified by their design, where positive indicates the homophily-based GNN performs better and negative indicates the heterophily-adjusted
    GNN performs better. The bars in each subplot represent the difference in F1 score between the models specified on the y-axis. Error bars denote the standard deviation in differences across runs. As the local homophily ratio greatly differs from the global homophily ratios, we find that models identified as designed for heterophily (GCNII, H2GNC) perform better.}
    \label{subset_synth_model_diff}
    \vspace{-0.4cm}
\end{figure}
\end{comment}

\noindent \textbf{(Q2) Performance discrepancies for homophilous vs. heterophilous GNNs.}
\begin{comment}
\begin{table}[]
\caption{Global F1 performance $\pm$ standard deviation across GNN models on synthetic data generated with global homophily ratios $h \in \{0.1, 0.3, 0.7, 0.9\}$ and uniformity $\rho = 0.5$. GCN performs slightly better in most settings, but performance is relatively uniform when accounting for variance. Bold denotes highest average F1 score.}
\vspace{-0.2cm}
\begin{tabular}{@{}llllll@{}}
\toprule
      & 0.1 & 0.3 & 0.7 & 0.9 \\ \midrule
GCN	& $ \mathbf{0.96 \pm 0.01 }$ & $ \mathbf{0.95 \pm 0.02 }$ &  $\mathbf{ 0.94 \pm 0.02} $ & $ 0.96 \pm 0.01 $\\
GAT	&$ 0.93 \pm 0.02 $ & $ 0.88 \pm 0.07 $ &  $ 0.88 \pm 0.05 $ & $ 0.93 \pm 0.03 $\\
SAGE	&$ 0.91 \pm 0.01 $ & $ 0.9 \pm 0.02 $ &  $ 0.89 \pm 0.01 $ & $ 0.92 \pm 0.02 $\\
GCNII	&$ 0.95 \pm 0.02 $ & $ 0.89 \pm 0.02 $ &  $ 0.87 \pm 0.02 $ & $ \mathbf{0.97 \pm 0.0 }$\\
H2GCN	&$ 0.93 \pm 0.01 $ & $ 0.89 \pm 0.02 $ &  $ 0.88 \pm 0.01 $ & $ 0.93 \pm 0.01 $\\ \bottomrule
\end{tabular}
\vspace{-.2cm}
\label{global}
\end{table}
\end{comment}
%\vspace{0.15cm}
%We now seek to answer whether GNNs built specifically for heterophily display different performance across varying local homophily ranges as compared to simpler GNN architectures.
To understand how different GNN designs amplify or reduce discrepancy, 
we analyze the trend of local performances across models. 
%All models achieve a comparable performance globally of 0.9-0.95 F1-score. 
While all models achieve similar global performance, 
%argues that the synthetic setup does not necessarily benefit from heterophilous GNNs, 
we observe that the models perform differently depending on the local homophily range. 
%Specifically, if a particular homophily range at test time has few nodes relative to the size of the graph, a large difference in performance between models may have minimal impact on the global metric. To assess this, we look at the difference in performance across pairs of models, particularly comparing those which utilize homophily and those that are designed for heterophily. 
%We focus on the extreme cases of $h \in \{0.1, 0.9\}$ and use a smaller set of bins relative to Fig.~\ref{fig:local_vs_global_homophily} to minimize variance due to insufficient nodes in various regions. Plots for $h \in \{0.3, 0.5, 0.7\}$ can be found in App.~\ref{app:synth_extra_comparisons}, but demonstrate minimal difference between the architectures.}
As shown in Figure \ref{fig:local_vs_global_homophily} (and Figure \ref{fig:pairwise_diff_synth} in App. \ref{app:synth_extra}), homophilous models often have higher performance for test nodes with local homophily levels that are close to the global homophily, \rebuttal{as expected from Theorem 1}, whereas heterophily-based models perform better than homophilous models for nodes with local homophily levels far from the global homophily, \rebuttal{agreeing with Theorem 2}. These insights highlight the different behaviors of the two designs, indicating that \textbf{models designed for heterophily are able to alleviate performance discrepancies across nodes, while homophilous models exacerbate discrepancy in strongly homophilous/heterophilous graphs}, \rebuttal{ Interestingly, LINKX, despite being MLP based, is competitive as compared to the other heterophilous GNNs in minimizing discrepancy. We attribute this to the separation of ego-node and neighbor-node embeddings, a key design in the other heterophilous models \cite{zhu2020beyond, platonov2023critical}.} These results verify that heterophilous models offer a better performance trade-off between nodes with over- and under-represented local homophily levels, displaying minor degradation on the over-represented nodes and significant improvement on the under-represented nodes. 
%Furthermore, the results help to contextualize the similar global results -- heterophilous models tend to perform  minority subgroups, 

%models such as GCN and GAT perform better on nodes with local homophily ratios similar to the global homophily ratio, but fail to generalize across the entire local homophily range, resulting in significant performance discrepancies.

\begin{comment}
\begin{table}[t]
\caption{Global F1 performance $\pm$ standard deviation across GNN models on real-world data. In heterophilous datasets, MLP dominates performance while on homophilous datasets many of the models perform equally well.}
\resizebox{0.8\columnwidth}{!}
{
\begin{tabular}{@{}lllllll@{}}
\toprule
 & Cornell & Wisconsin & Squirrel  & Cora & Coauthor - CS \\ \midrule
MLP& $\mathbf{0.78  \pm  0.06}$&$\mathbf{0.85  \pm  0.03}$&$\mathbf{0.4  \pm  0.03}$&$0.68  \pm  0.03$&$0.79  \pm  0.03$\\
GCN& $0.58  \pm  0.1$&$0.55  \pm  0.07$&$0.29  \pm  0.02$&$\mathbf{0.86  \pm  0.0}$&$0.91  \pm  0.02$\\
GAT& $0.6  \pm  0.08$&$0.57  \pm  0.11$&$0.29  \pm  0.04$&$0.85  \pm  0.01$&$0.84  \pm  0.01$\\
SAGE& $0.54  \pm  0.07$&$0.54  \pm  0.08$&$0.35  \pm  0.03$&$0.84  \pm  0.01$&$0.92  \pm  0.0$\\
GCNII& $0.56  \pm  0.1$&$0.41  \pm  0.15$&$0.32  \pm  0.07$&$0.75  \pm  0.02$&$0.82  \pm  0.0$\\
H2GCN& $0.7  \pm  0.07$&$0.77  \pm  0.06$&$0.34  \pm  0.01$&$\mathbf{0.86  \pm  0.0}$&$\mathbf{0.93  \pm  0.0}$\\\bottomrule
\end{tabular}
}
\end{table}
\end{comment}
\section{Real-world Empirical Evaluation}
\label{real_section}

We now demonstrate how our results extend to real-world datasets.
%We begin by analyzing homophilous datasets and characterizing their performance across different local homophily ranges, then compare their behavior to the heterophilous datasets. 
%We note that our findings from a global perspective agree with those in the literature \cite{zhu2020beyond, ma2021homophily, du2022gbk}, while additionally provide a deeper understanding to their performance from a local perspective. 

\noindent \textbf{Data and Setup.} We choose five real-world graphs: two  homophilous (Cora \cite{sen2008}, Coauthor-CS 
\cite{bojchevski2017deep}) and three heterophilous (Wisconsin \cite{Pei2020Geom}, Squirrel \cite{rozemberczki2021multi}, \rebuttal{and Arxiv-Year \cite{lim2021large})}. \rebuttal{Each heterophilous dataset varies greatly in size, providing an opportunity to measure the impact of the number of nodes and density on discrepancy}. Additionally, we choose Wisconsin and Squirrel due to their historically inconclusive performance when comparing GNNs to non-graph-based deep learning baselines~\cite{zhu2020beyond, ma2021homophily}. Our analysis aims to provide insight into why prior results have been inconclusive and explain their poor performance from a local perspective. For Wisconsin and Squirrel, we perform 30 random 50-25-25\% splits of the nodes to obtain the train, validation, and test sets. For Arxiv-Year, Cora, and Coauthor-CS we perform 5 random splits over the same ratio as the graph are much larger. 
%\vspace{0.1cm}
%\noindent \textbf{Evaluation.} 
We train the same architectures as in the synthetic experiments, seen in Section 5.2. 

We report results in Figure \ref{real_world} of the main text, and Figure \ref{fig:real_extras} in App. \ref{app:real_world_extra}. While we group all test nodes into four local homophily ranges, we limit the ranges of Wisconsin in Figure \ref{real_world} due to having less than three test nodes in local homophily ranges above 0.6. Additionally, features may be more informative in certain local homophily ranges, obscuring when performance degrades due to homophily or uninformative features. Thus, we report the difference in F1 score between the different GNNs and the graph-agnostic MLP, $\Delta$F1 $\equiv$ (F1-score GNN - F1-score MLP), to disentangle how performance relies on the node features as compared to the graph structure.

%While the synthetic experiments were designed to ensure the neighborhood structure was required to correctly predict the node label, we cannot easily enforce that on our real-world datasets. Thus, to aid in our analysis, we also include a simple MLP model which relies solely on the node features to make predictions. By comparing the MLP to the GNN models, it becomes possible to disentangle how performance relies on the node features as compared to the graph structure. 

\subsection{\textbf{Performance Discrepancies in Homophilous vs. Heterophilous Real-World Datasets}}

\begin{figure*}[t]
    \centering
     \includegraphics[width=\textwidth, keepaspectratio]{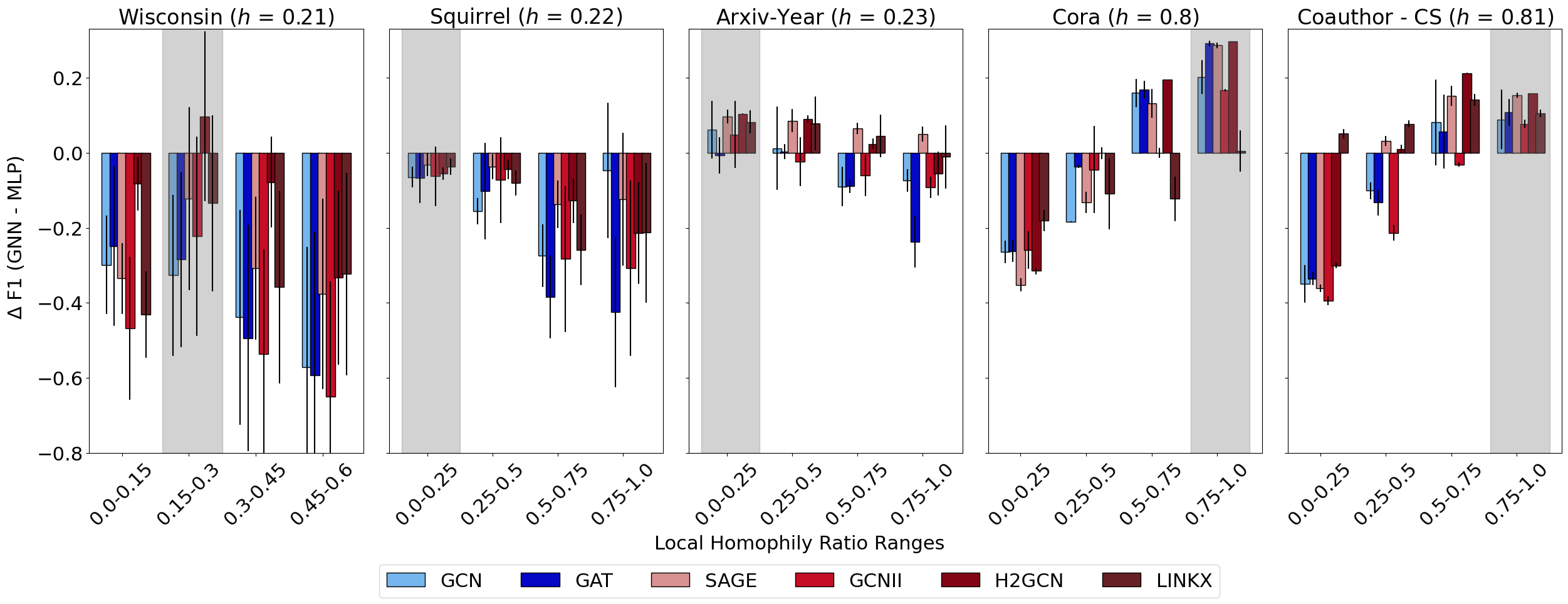}

     \caption{Real-words graphs: Difference in performance, $\Delta$F1=F1(GNN) - F1(MLP), for GNN models across ranges of local homophily ratios (more GNNs  in App. \ref{app:real_world_extra}), averaged over  multiple splits. 
     Our results elucidate \textit{how} models achieve different global performance, where heterophilous models (bars in reds) better tend to combat the systematic performance discrepancy seen in homophilous models (bars in blues).
     % For a range specified on the x-axis, the bars denote $\Delta$F1 for each GNN across test nodes that fall within the local homophily range. 
     Gray indicates the range that the global homophily ratio falls in; negative bars indicate worse performance than MLP. 
     %, and %, while positive bars indicate better performance. 
     % error bars are the standard deviation in performance across experiment runs. %Blues are used for GNNs with homophilous designs, while reds for GNNs with heterophilous designs.
     }
     \label{real_world}
% \vspace{-.15cm}
\end{figure*}

\noindent \textbf{Homophilous Graphs.} The homophilous graphs are shown in the two right-most plots of Figure \ref{real_world}. First, we highlight the nearly \textbf{0.6 drop in $\Delta$F1 across both datasets for test nodes with local homophily ratios far from the global homophily ratio}, performing worse than an MLP as denoted by the negative F1 score difference. Furthermore, this degradation is consistent across all of the GNN architectures, with H2GCN, which leverages heterophilous designs, maintaining generally higher performance as the nodes become more heterophilous. This result demonstrates the practical implications of our theoretical analysis, allowing us to additionally demonstrate that degradation can occur under rich feature sets. Notably, the MLP outperforms all of the GNN architectures in the heterophilous parts of the graph, implying the neighborhood information is actively corrupting performance in these regions due to the GNN's reliance on the global homophily. 

\noindent \textbf{Heterophilous Graphs.} The heterophilous datasets are shown on the three left-most plots of Figure \ref{real_world}. Compared to the homophilous datasets, Wisconsin and Squirrel have poor performance relative to the MLP for nearly all local homophily ranges. Our local analysis identifies a mechanism for \textit{how} this arises: exacerbated performance degradation for nodes with local homophily ratios far from the global homophily ratio.
%For instance, when comparing the real-world heterophilous datasets to our synthetic heterophilous datasets, models can perform exceptionally well despite the heterophilous setting when the features and graph structure are causally linked. Future work providing deeper analysis on the relationship between feature generation and structure in these heterophilous datasets could help elucidate why removing the graph structure is beneficial for learning. }
Despite the MLP model outperforming the GNN models across Wisconsin and Squirrel, there is still a notable degradation in performance as the local homophily ranges increase, \textbf{causing a 0.2-0.4 drop in $\Delta$F1 between the bins furthest from the global homophily ratio}. We hypothesize that degree can influence whether GNNs degrade in performance on heterophilous graphs, as seen in our theoretical analysis in App. \ref{app:proof}, causing this drop to be lower as compared to the homophilous graphs. \rebuttal{Aligning closer to the homophilous datasets, each model's best performance on Arxiv-Year is within the bin closest to the global homophily ratio, demonstrating a scenario where the heterophilous neighborhood is beneficial to learning.} Following the trends seen in the synthetic data, the heterophilous models also display higher uniformity across the homophily levels for each datasets. Surprisingly, for Squirrel, we observe a significant difference in MLP performance across the various local homophily ranges, indicating that the features are intrinsically more informative in certain regions. When accounting for this through $\Delta$F1, we see similar degradation to that in the synthetic and homophilous datasets, again with the heterophilous GNNs tending to perform best.

\subsection{\textbf{Discussion}}

Two key insights emerge that have yet to be established in other analyses: (a) nodes with higher local homophily are not inherently easier to classify, as seen by the relative drops in performance on heterophilous datasets, and (b) heterophilous designs improve learning across nearly all local homophily ranges, not just one particular range, alleviating performance discrepancies across nodes. 
%It is important to note that our theoretical and synthetic analyses do not take into account varying levels noise in node feature across the graph, presenting an opportunity to further disentangle the signal from features and local homophily in a systematic analysis. 
%That said, results on Cora and Coauthor - CS present scenarios where GNN performance degrades drastically, potentially indicating a scenario where despite informative features (as indicated by MLP performance), the GNNs that are biased towards homophily are unable to make use of the heterophilous connections. 
We note that there might be additional factors which give rise to performance variations across local homophily ratios. For instance, previous works have identified that, under certain settings, nodes with extreme homophily levels can be easier to classify \cite{ma2021homophily, du2022gbk, cavallo20222}. Moreover, the ease of classification has been tied to the interaction of degree and homophily, pointing towards high degree nodes as the easiest setting \cite{yan2021two}. Performance variations due to raw structural properties (e.g. degree or local homophily ratio) and local homophily shift are not necessarily mutually exclusive, nor do they conflict with our results. Instead, we hypothesize both factors can interact, amplifying performance disparities across homophily ranges, further necessitating future work which studies their interplay.

\section{Conclusion}
In this work, we take a local perspective focused on discerning how nodes of varying local homophily levels can experience performance discrepancies. We first theoretically demonstrated that classification performance degrades as the local homophily ratio of a node deviates from the global homophily ratio. To demonstrate the generalizability of our findings, we proposed a new parsimonious synthetic graph generator that allows generating graphs with varying global and local homophily. We demonstrated that our theoretical insights still hold in more general settings, finding that performance degradation can occur in either highly homophilous or heterophilous settings. Furthermore, we showed that this disparity in performance can be reduced by using GNN models which adopt explicit mechanisms to support heterophily, providing insight into how GNNs with heterophilous designs improve performance globally. Our experiments on real-world datasets of varying global homophily ratios confirm the practical implications of our insights, exhibiting similar disparity patterns. The discovery and characterization of GNN degradation through shifts in local homophily relative to a graph's global homophily necessitates the development of new GNNs that are able to explicitly handle such data shifts. Additionally, our findings highlight how, in human-facing applications of GNNs, individuals might experience disparate treatment under a GNN due to structural properties of the underlying graph, opening new research directions in algorithmic fairness. 

\bibliographystyle{unsrtnat}
\bibliography{reference}
% For bibLaTeX users:
% \printbibliography

\appendix
\clearpage

\appendix
\section{Appendix}

\subsection{Notation Table}
\label{app:notation}

In this section, we give a summary of the notation used throughout the main text. 

\begin{table}[h]
    \centering
    \caption{Graph- and Node-level Notations and Definitions}
    \label{tab:graph}
    \begin{tabularx}{\textwidth}{p{30mm}X}
    \toprule
    \textbf{Notation} & \textbf{Definitions} \\
    \midrule
    $G$ & Graph $G$ \\
    $V$ & Node set $V$ for $G$ \\
    $E$ & Edge set $E$ for $G$ \\
    $\mathbf{I}$ & Identity matrix \\ 
    $\mathbf{A}$ & Adjacency matrix for $G$. Shape $|V| \times |V|$ \\
    $\mathbf{X}$ & Feature matrix for $G$. Shape $|V| \times f$ where $f$ is the number of features. \\
    $\mathbf{Y}$ & One-hot encoded class matrix for $G$. Shape $|V| \times c$ where $c$ is the number of classes. \\
    $\mathbf{H}_{L}$ & Class compatibility matrix..  \\
    $h$ & Global homophily ratio. \\
    \midrule
    
    $i$ & A node $i \in G$ \\  
    $\mathbf{x_{i}}$ & Feature vector for node $i$ \\
    $y_{i}$ & Class label for node $i$ \\
    $\mathbf{y_{i}}$ & One-hot encoded class label vector for node $i$ \\
    $N_{k}(i)$ & $k$-hop neighborhood for a node $i$ \\
    $h_{i}$ & Local homophily ratio for a node $i$ \\
    $\alpha_{i}$ & Shift in local homophily ratio for a node $i$, relative to the global homophily ratio of its graph, $h + \alpha_{i} = h_{i}.$ \\
    \bottomrule
    \end{tabularx}
    
\end{table}

\vspace{-0.5cm}
\begin{table}[h]
    \centering
    \caption{Graph Neural Network Notations and Definitions}
    \label{tab:gnn}
    \begin{tabularx}{\textwidth}{p{30mm}X}
    \toprule
    \textbf{Notation} & \textbf{Definitions} \\
    \midrule
    $n_{train}$ & Training nodes of a GNN \\
    $\mathbf{W_{l}}$ & The weight matrix for a layer $l$ of a GNN \\
    $\mathbf{R_{l}}$ & The hidden representations for nodes at layer $l$ of a GNN. $R_{0}$ denotes the input features. \\
    $\mathbf{r_{i}^{l}}$ & The hidden representation vector for a node $i$ at layer $l$ \\
    $\sigma$ & Activation function of a GNN. \\
    $\mathbf{z_{i}}$ & Output logit vector of a GNN for node $i$. \\
    \bottomrule
    \end{tabularx}
\end{table}

\vspace{-0.5cm}
\begin{table}[h]
    \centering
    \caption{Synthetic Graph/Evaluation Notations and Definitions}
    \label{tab:eval}
    \begin{tabularx}{\textwidth}{p{30mm}X}
    \toprule
    \textbf{Notation} & \textbf{Definitions} \\
    \midrule
    $n$ & Number of nodes \\
    $m$ & Number of edges added per generation step \\
    $\epsilon$ & Noise strength for generated features. \\
    $\rho$ & Probability to sample a new homophily ratio. \\
    $\delta$ & Hyperparameter to determine if a node has drifted in homophily level. \\
    $\Delta F1$ & F1 Score of GNN - F1 Score of MLP (graph agnostic). \\
    \bottomrule
    \end{tabularx}
\end{table}

\subsection{Theoretical Relationship between a Node's Local Homophily Level and Performance}

In this section, we provide the proof for Theorem 1 in the main text. We additionally perform theoretical analysis on the influence of degree on Theorem 1, as well as the impact of higher order aggregation. We provide a high-level takeaway for each additional analysis.

\subsubsection{Proof for Theorem 1: Homophilous Model $\mathbf{(A+I)XW}$}
\label{app:proof}

As described in the main text, we consider a graph $G$ with a subset of nodes $n_{train}$. Each node $i \in n_{train}$ has an associated node feature vector $\mathbf{x_{i}}$, one-hot encoded class label vector $\mathbf{y_{i}}$, 1-hop homophily ratio $h$, and degree $d$. For brevity, we focus on binary classification, and represent $\mathbf{y_{i}}$ as $\begin{bmatrix} 1 & 0 \end{bmatrix}$ when $y = 0$ and $\begin{bmatrix} 0 & 1 \end{bmatrix}$ when $y = 1$. The node feature vectors are initialized as $\mathbf{x_{i}} = \begin{bmatrix} (0.5+p) & (0.5-p) \end{bmatrix}$ when $y_i = 0$ and $\mathbf{x_{i}} = \begin{bmatrix} (0.5-p) & (0.5+p) \end{bmatrix}$ when $y_{i} = 1$. The GNN, $F$, is formulated as $\mathbf{(A + I)XW}$, where $\mathbf{A + I}$ is $G$'s adjacency matrix with self-loops and $\mathbf{W}$ is $F$'s weight matrix, and trained through $n_{train}$. The final prediction for a node $i$ is $\mathbf{argmax} \, \mathbf{{z}_{i}}$ where $\mathbf{{z}_{i}}$ is the output logit vector of $F$. Focusing first on the $\textbf{(A + I)X}$ term, a row (node) $i$ in this matrix with a class 0 will be equal to: 

\begin{comment}
    $r_{i, 0}$ \ben{what's an updated representation? what does $r_{i, 0}$ represent?} as seen in equation \ref{z0}
\end{comment}

\begin{equation}
\begin{aligned}
y_{i} = 0: \mathbf{r_{i}^{1}} = \mathbf{x_{i}} + \dfrac{hd}{2} \begin{bmatrix} 1+p & 1-p \end{bmatrix} + \dfrac{(1-h)d}{2} \begin{bmatrix} 1-p & 1+p \end{bmatrix}
\end{aligned}
\label{z0}
\end{equation}
similarly, a row (node) $i$ in this matrix of class 1 will be equal to:

\begin{equation}
\begin{aligned}
y_{i} = 1: \mathbf{r_{i}^{1}} = \mathbf{x_{i}} + \dfrac{hd}{2} \begin{bmatrix} 1-p & 1+p \end{bmatrix} + \dfrac{(1-h)d}{2} \begin{bmatrix} 1+p & 1-p \end{bmatrix}
\end{aligned}
\label{z1}
\end{equation}
The computations for each $\mathbf{r_{i}^{1}}$ represent the sum of the node vector for node $i$, the feature vectors from the homophilous connections of node $i$, and the feature vectors of the heterophilous connections for node $i$. Condensing these terms, the simplified expressions for each $\mathbf{r_{i}^{1}}$ is shown in Equations \ref{simple_z0} and \ref{simple_z1}

\begin{equation}
\begin{aligned}
y_{i} = 0: \mathbf{r_{i}^{1}} = \dfrac{1}{2} \begin{bmatrix} (1+d+p+dp(2h-1)) & (1-p+d(1+p-2hp)) \end{bmatrix} 
\end{aligned}
\label{simple_z0}
\end{equation}

\begin{equation}
\begin{aligned}
y_{i} = 1: \mathbf{r_{i}^{1}} = \dfrac{1}{2} \begin{bmatrix} (1-p+d(1+p-2hp)) & (1+d+p+dp(2h-1)) \end{bmatrix} 
\end{aligned}
\label{simple_z1}
\end{equation}

We can then represent $\textbf{(A+I)XW}$ as $\textbf{RW} $ as seen in Equation \ref{zw} and set it equal to the label matrix $\mathbf{Y}$ after a transformation by the weight matrix $\textbf{W}$. 

\begin{equation}
\begin{aligned}
\textbf{R} = \dfrac{1}{2} \left[ {\begin{array}{cc}
                                    \vdots & \vdots \\
                                    (1+d+p+dp(2h-1)) & (1-p+d(1+p-2hp)) \\
                                    \vdots & \vdots \\
                                    (1-p+d(1+p-2hp)) & (1+d+p+dp(2h-1)) \\
                                    \vdots & \vdots \\
                                    \end{array} } \right]_{|T| \times 2}
\end{aligned}
\label{zw}
\end{equation}

\begin{equation}
\begin{aligned}
\\ \textbf{R}_{|T| \times 2}\textbf{W}_{2 \times 2} = \textbf{Y}_{|T| \times 2} = \begin{bmatrix}
                                                        \vdots & \vdots \\
                                                       1 & 0 \\
                                                       \vdots & \vdots\\
                                                       0 & 1 \\
                                                       \vdots & \vdots\\
                                                       \end{bmatrix}
\end{aligned}
\label{zw_y}
\end{equation}

Our goal is to now solve for the optimal $\textbf{W}$ in this system of equations. While the system of equations is overdetermined, each row of the same class share the same solution. Thus, we can sample the unique data points, leaving us with the optimal $\textbf{W}$ in Equation \ref{opt_w} from solving the reduced system of equations, where $c_{1} = \dfrac{1}{2(p-d^{2}p + 2dhp + 2d^{2}hp)}$.

\begin{equation}
\begin{aligned}
\textbf{W} = c_{1} \left[ {\begin{array}{cc}
                                    1 + d + p + dp(2h - 1) & -1 + p - d(1 + p - 2hp) \\
                                    -1 + p - d(1 + p - 2hp) & 1 + d + p + dp(2h - 1) \\
                                    \end{array} } \right] 
\end{aligned}
\label{opt_w}
\end{equation}

We now consider a new test point, $t$ with label $y_{t} = 0$ and subsequently features $\mathbf{x_{t}} = \dfrac{1}{2} \begin{bmatrix} (1+p) & (1-p) \end{bmatrix}$. In addition, we assume that $t$ has a homophily ratio $h + \alpha_{t} = h_{t} \in [0, 1]$ where $\alpha_{t}$ represents a shift away from $h$. When $\alpha_{t}$ is positive, we can interpret this as a test node with a higher homophily ratio, and when $\alpha_{t}$ is negative, a lower homophily ratio. Under this new homophily ratio, $\mathbf{r_{t}^{1}}$ can be calculated as:

\begin{equation}
\begin{aligned}
\mathbf{r_{t}^{1}} = \dfrac{h_{t}d + 1}{2} \begin{bmatrix} 1+p & 1-p \end{bmatrix} + \dfrac{(1-h_{t})d}{2} \begin{bmatrix} 1-p & 1+p \end{bmatrix}
\end{aligned}
\label{zt}
\end{equation}

Similar to equation \ref{simple_z0}, the simplified expressions for $\mathbf{r_{t}^{1}}$ is given as:

\begin{equation}
\begin{aligned}
\mathbf{r_{t}^{1}} = \dfrac{1}{2} \begin{bmatrix} (1+d+p+dp(2h_{t}-1)) & (1-p+d(1+p-2h_{t}p))\end{bmatrix}
\end{aligned}
\label{simple_zt}
\end{equation}

We can now compute $\mathbf{r_{t}^{1}} \textbf{W}$ and analyze the associated predictions. The output of $\mathbf{r_{t}^{1}} \textbf{W}$ is:

\begin{equation}
\mathbf{z_{t}} = \dfrac{1}{1 +d(2h-1)} \begin{bmatrix} (1+d(h+h_{t}-1)) & d(h-h_{t})\end{bmatrix}
\label{equation_ztw}
\end{equation}

It is easy to check that when $h_{t} = h$, we recover the correct prediction of $[1, 0]$ as expected. We now consider what happens when $h_{t} = h + \alpha_{t}$ where $\alpha_{t} \in [-h, 1-h]$. Since we are interested in understanding how the predictions would change as a byproduct of $\alpha_{t}$, we look at the difference between predictions when $h_{t} = h$. Below we compute $\Delta \mathbf{z_{t}}$, the change in the logit vector output for a test point with local homophily $h_{t}$, where $b_{1} = \dfrac{d}{1 + d(2h-1)}$.

\begin{align}
\Delta  \mathbf{z_{t}}  &= 
\dfrac{b_{1}}{d} 
\begin{bmatrix} 
(1+d(h+(h+\alpha_{t})-1)) & d(h-(h+\alpha_{t}))
\end{bmatrix}  \nonumber \\
&-\dfrac{b_{1}}{d} \begin{bmatrix} (1+d(h+h-1)) & d(h-h)\end{bmatrix}\\
%\Delta  \mathbf{z} &= \dfrac{b_{1}}{d}  \begin{bmatrix} (1+2dh+d\alpha_{t}-d - 1-2dh+d) & -(d\alpha_{t}) \end{bmatrix} \\
\Delta \mathbf{z_{t}} &= b_{1} \begin{bmatrix} \alpha_{t} & -\alpha_{t} \end{bmatrix} \quad \square
\label{equation_ztw_simplified}
\end{align}

%%%%%%%% HETEROPHILOUS MODEL
\rebuttal{\subsubsection{Proof for Theorem 2: Heterophilous Model $\mathbf{(X \mathbin\Vert AX)W}$}
\label{app:hetero_proof}
Following the same set up previously described for the GNN parameterized as $\mathbf{(A+I)XW}$, we also consider a GNN, $F'$, parameterized as $\mathbf{(X \mathbin\Vert AX)W}$, concatenating the features from the ego-node and the aggregated neighboring features. This separation through concatenation is found in a variety of GNNs, and is often denoted as a core design to promote heterophilous learning \cite{hamilton2017sage, zhu2020beyond}.}  

\rebuttal{Focusing first on the aggregation step of $F'$, $\mathbf{AX}$, a node $i$ of class $0$ will have aggregated features from its neighbors, $\mathbf{r_{i}^{1}}$, defined as: 
\begin{equation}
\begin{aligned}
y_{i} = 0: \mathbf{r_{i}^{1}} =  \dfrac{hd}{2} \begin{bmatrix} 1+p & 1-p \end{bmatrix} + \dfrac{(1-h)d}{2} \begin{bmatrix} 1-p & 1+p \end{bmatrix}
\end{aligned}
\label{z0_ax}
\end{equation}
similarly, a node $i$ of class 1 will have aggregated features from its neighbors:
\begin{equation}
\begin{aligned}
y_{i} = 1: \mathbf{r_{i}^{1}} =  \dfrac{hd}{2} \begin{bmatrix} 1-p & 1+p \end{bmatrix} + \dfrac{(1-h)d}{2} \begin{bmatrix} 1+p & 1-p \end{bmatrix}
\end{aligned}
\label{z1_ax}
\end{equation}
Then, the concatenated vectors $\mathbf{(X \mathbin\Vert AX)W}$ for nodes of class 0 and 1 are expressed as: 
\begin{equation}
\begin{aligned}
y_{i} = 0: \mathbf{x_{i} \mathbin\Vert r_{i}^{1}} =   \dfrac{1}{2}
    \begin{bmatrix} 1+p & 1-p & d + dp (2h - 1) & d(1 + p - 2 h p)\end{bmatrix} 
\end{aligned}
\label{z0_concat}
\end{equation}
\begin{equation}
\begin{aligned}
y_{i} = 1:  \mathbf{x_{i} \mathbin\Vert r_{i}^{1}} = \dfrac{1}{2}\begin{bmatrix} 1 - p & 1 + p & d (1 + p - 2 h p) & d + dp (2h - 1)\end{bmatrix} 
\end{aligned}
\label{z1_concat}
\end{equation}
We refer to the collection of concatenated vectors as $\mathbf{R} \in \mathcal{R}^{|V| \times 4}$. Then, the goal, similar to the case of the GNN $F$, is to solve for $W$ such that:
\begin{equation}
\begin{aligned}
\\ \textbf{R}_{|T| \times 4}\textbf{W}_{4 \times 2} = \textbf{Y}_{|T| \times 2} = \begin{bmatrix}
                                                        \vdots & \vdots \\
                                                       1 & 0 \\
                                                       \vdots & \vdots\\
                                                       0 & 1 \\
                                                       \vdots & \vdots\\
                                                       \end{bmatrix}
\end{aligned}
\label{zw_y_concat}
\end{equation}
Following a similar procedure to the analysis of $F$, we reduce the system of equations by sampling the unique points and solve for the optimal $\mathbf{W}$. However, unlike in $F$, $\mathbf{R}$ is not a square matrix after sampling, and thus we use a right pseudo-inverse through $\mathbf{R^{-1}= R^T (RR^{T})^{-1}}$ to attain a solution to the system of equations.}  

\rebuttal{We now consider a new test point, $t$ with label $y_{t} = 0$, similar to the set up for $F$. Then:
\begin{equation}
\begin{aligned}
\mathbf{x_{t} \mathbin\Vert r_{t}^{1}} =   \dfrac{1}{2}
    \begin{bmatrix} 1+p & 1-p & d + dp (2h_{t} - 1) & d(1 + p - 2 h_{t} p)\end{bmatrix} 
\end{aligned}
\label{z0_concat_test}
\end{equation}
We can now compute $\mathbf{r_{t}^{1}}\mathbf{W} = \mathbf{z_{t}}$: 
\begin{equation}
\begin{aligned}
\mathbf{z_{t}} =   
    \dfrac{1}{1 + d^2 (2h-1)^2}\begin{bmatrix} 1 + d^2 (2h - 1) (h + h_{t} - 1)) & (d^2 (2h - 1) (h - h_{t})) \end{bmatrix} 
\end{aligned}
\label{zt_concat}
\end{equation}
As before, when $h_{t} = h$, it is easy to see that the label is recovered. We now consider when $h_{t} = h + \alpha_{t}$ and look at the difference in predictions relative to when $h_{t} = h$. $\Delta \mathbf{z_{t}}$ for a test point with local homophily $h_{t}$ predicted by the heterophilous model, where $b'_{1} = \dfrac{d^2 (2h - 1)}{1 + d^2(2h - 1)^{2}}$ is: 
\begin{align}
 \Delta \mathbf{z_{t}} &= b'_{1} \begin{bmatrix} \alpha_{t} & -\alpha_{t} \end{bmatrix} \quad \square
\label{equation_ztw_simplified_concat}
\end{align}
Thus, we can still expect heterophilous models which adopt the ego-neighbor separation design to degrade in performance as the local homophily of a node changes, just as studied in the Section \ref{theory_section} of the main text. However, the change in coefficient, $b_{1}$ vs. $b'_{1}$, demonstrates a significantly different degradation pattern.}

\begin{figure}[h]
    \centering
    \includegraphics[width=0.6\textwidth, keepaspectratio]{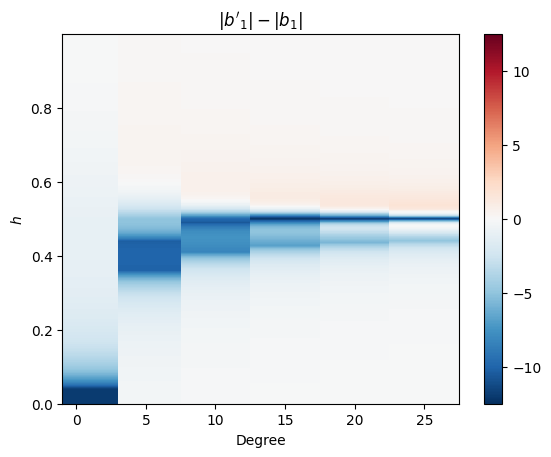}
    \caption{\rebuttal{The difference in magnitudes for $b'_{1}$ and $b_{1}$ for parameter combinations of degree and $h$. Red indicates parameter combinations where $b_{1}$, coming from the homophilous model, is smaller, and blue indicates parameter combinations where $b'_{1}$, coming from the heterophilous model, is smaller. Smaller coefficients are preferred to minimize the discrepancies on predictions induced by $\alpha_t$. Predominantly negative values indicates that the heterophilous model formulated as $\mathbf{(X \mathbin\Vert AX)W}$ is less sensitive to $\alpha_t$ and thus produces smaller discrepancies.}}
    \label{diff_coeff}
\end{figure}

\rebuttal{\noindent \textbf{The impact of $b_{1}$ and $b'_{1}$.} We analyze how the different coefficients influence the degradation of performance. To do this, we compare the difference in magnitude between $b'_{1}$ and $b_{1}$, where a positive difference indicates a smaller magnitude for $b_{1}$ (indicating that $b'_{1}$ has a stronger influence on $\Delta \mathbf{z_{t}}$ as compared to $b_{1}$), while a negative difference indicates a smaller magnitude for $b'_{1}$ (indicating that $b_{1}$ has a stronger influence on $\Delta \mathbf{z_{t}}$ as compared to $b'_{1}$). As the goal is to identify which GNN formulation is likely to produce discrepancies, we prefer models with smaller magnitudes given this will minimize the influence of $\alpha_{t}$.}

\rebuttal{The comparison of $b'_{1}$ and $b_{1}$ is performed in Figure \ref{diff_coeff}, where we sample points for the degree from $\{1, 5, 10, 15, 20, 25\}$ and $h$ from $0$ to $1$ in intervals of $0.001$. We can see that the difference in the coefficients is predominantly negative, indicating that the GNN $F'$, formulated as $\mathbf{(X \mathbin\Vert AX)W}$, is less impacted by shifts in local homophily. Thus, although both the homophilous and heterophilous formulations are impacted by discrepancies induced by $\alpha_{t}$, we can expect the predictions of $F'$ to experience less discrepancy as the magnitude of $\alpha_{t}$ becomes larger.}

\subsubsection{The Impact of Degree on $\Delta \mathbf{z_{t}}$ on Homophilous Models ($\mathbf{AXW}$)}

As in the main text, we provide an outline of the cases below, with commentary on the impact of degree $d$ on the cases. 

\noindent \textbf{Setting 1: Heterophilous $(\mathbf{0 \le \textit{h} < 0.5}$)}: In this scenario, when $d(2h - 1) < -1$, $b_{1} < 0$, leading to $\mathbf{z_{t}} =  \mathbf{y_{t, 0}} + |b_{1}|\begin{bmatrix} -\alpha_{t} & \alpha_{t} \end{bmatrix}$ where the sign of $b_{1}$ has been distributed into the vector. There are then two cases: 
    \begin{equation}
    \mathbf{z_{t}} = 
        \begin{cases}
          \mathbf{y_{t}} + |b_{1}|\begin{bmatrix} |\alpha_{t}| & -|\alpha_{t}| \end{bmatrix},  & \text{if}\ h_{t} \le h \\
          \mathbf{y_{t}} + |b_{1}|\begin{bmatrix} -\alpha_{t} & \alpha_{t} \end{bmatrix},  & \text{if}\ h_{t} > h .
        \end{cases}
    \label{case1}
    \end{equation}

where the sign of $\alpha_{t}$ has been integrated into the vectors. To better understand the conditions that lead to this scenario, we note that $d$ must be greater than $\dfrac{1}{1-2h}$ to lead to $b_{1} < 0$. However, even for moderate values of $h$, the required degree to elicit a change is small, thus the heterophily shift is often what will dominate the prediction, not the degree. Furthermore, as $d(2h - 1)$ becomes more negative, $b_{1}$ will tend towards $\dfrac{1}{1 + (2h - 1)}$, demonstrating that the influence of $d$ on predictions decreases as $d$ gets larger. As such, we can expect changes in the local homophily to be the dominant factor in prediction changes. 

\vspace{0.1cm}
\noindent \textbf{Setting 2: Mixed Homophily ($\mathbf{\textit{h} = 0.5}$)}: When the graph is not strongly homophilous nor strongly heterophilous, $h = 0.5 \rightarrow b_{1} = d$, leading to the following changes for $\mathbf{z_{t}}$:
    \begin{equation}
    \mathbf{z_{t}} = 
        \begin{cases}
          \mathbf{y_{t}} + d\begin{bmatrix} -|\alpha_{t}| & |\alpha_{t}| \end{bmatrix},  & \text{if}\ h_{t} \le h  \\
          \mathbf{y_{t}} + d\begin{bmatrix} \alpha_{t} & -\alpha_{t} \end{bmatrix}, & \text{if}\ h_{t} > h .
        \end{cases}
    \end{equation}

As the exception from other cases, the change in performance directly depends on the degree $d$ and thus performance variations may be dominated by degree, rather than local homophily, as $d$ becomes large. Intuitively, if the global homophily level is 0.5, we can expect the model to not make use of this signal.

\vspace{0.1cm}
\noindent \textbf{Setting 3: Homophilous ($\mathbf{0.5 < \textit{h} \le 1}$)}: In this scenario, $b_{1} > 0$, thus:
\begin{equation}
    \mathbf{z_{t}} = 
        \begin{cases}
          \mathbf{y_{t}} + |b_{1}|\begin{bmatrix} -|\alpha_{t}| & |\alpha_{t}| \end{bmatrix},  & \text{if}\ h_{t} \le h \\
          \mathbf{y_{t}} + |b_{1}|\begin{bmatrix} \alpha_{t} & -\alpha_{t} \end{bmatrix},  & \text{if}\ h_{t} > h .
        \end{cases}
    \label{case3}
    \end{equation}

Similar to Case 1, as $d(2h - 1)$ becomes more positive, $b_{1}$ will approach $\dfrac{1}{1+(2h - 1)}$, leading to $d$ becoming less influential on the prediction. As such, we can expect changes in the local homophily to be the dominant factor in prediction changes. 

\noindent \textbf{High-level Takeaway:} Notably, the heterophilous and homophilous settings are not necessarily symmetric -- the heterophilous setting requires the degree criteria to be satisfied in order to attain the predictive pattern, while the homophilous setting is always true. We suspect this plays a role in our (as well as previous) analysis as our theoretical analysis suggests homophilous datasets will always degrade in performance as test nodes become heterophilous, while heterophilous datasets will only degrade when the degree criteria is met. This phenomenon is seen in Figure \ref{real_world} of our real-world analysis, where the performance disparities are generally more drastic in the homophilous datasets.

\subsubsection{Additional Theoretical Analysis on 2-layer GNN}

\label{app:2_layer_proof}

We follow a similar set up as in the 1-layer case for the graph $G$, but now consider a 2-layer GNN, $F$, formulated as $\mathbf{(A + I)^{2}XW}$. 
During the first aggregation step of $F$, a node $i$ of class $0$ will have updated features as specified in Equation \ref{z0}, and a node $i$ of class $1$ will have updated features as specified in Equation \ref{z1}.

During the second aggregation step of $F$, a node $i$ of class 0 will have updated features $\mathbf{r_{i}^{2}}$ defined as: 
\begin{equation}
\begin{aligned}
y_{i} = 0: \mathbf{r_{i}^{2}} =  \dfrac{hd + 1}{2} \begin{bmatrix} (1+d+p+dp(2h-1)) & (1-p+d(1+p-2hp)) \end{bmatrix} + \\ \dfrac{(1-h)d}{2} \begin{bmatrix} (1-p+d(1+p-2hp)) & (1+d+p+dp(2h-1)) \end{bmatrix} 
\end{aligned}
\label{z02}
\end{equation}

similarly, a node $i$ of class 1 will have updated features:

\begin{equation}
\begin{aligned}
y_{i} = 1: \mathbf{r_{i}^{2}} =  \dfrac{hd + 1}{2} \begin{bmatrix} (1-p+d(1+p-2hp)) & (1+d+p+dp(2h-1)) \end{bmatrix}  + \\ \dfrac{(1-h)d}{2} \begin{bmatrix} (1+d+p+dp(2h-1)) & (1-p+d(1+p-2hp)) \end{bmatrix} 
\end{aligned}
\label{z12}
\end{equation}

Each $\mathbf{r_{i}^{2}}$ is the sum of the updated node vector for node $i$ after one aggregation step ($\mathbf{r_{i}^{1}}$), the updated feature vectors from the homophilous connections of node $i$, and the updated feature vectors of the heterophilous connections for node $i$. Condensing these terms, the simplified expressions for each $\mathbf{r_{i}^{2}}$ is shown in Equations \ref{simple_z02} and \ref{simple_z12}

\begin{equation}
\begin{aligned}
y_{i} = 0: \mathbf{r_{i}^{2}} = \dfrac{1}{2}
\left[\begin{matrix}
  1 + p + d^2 (1 + (1 - 2 h)^2 p) + d (2 + (4h - 2) p),\\
\end{matrix}\right.\\
\qquad\qquad
\left.\begin{matrix}
   1 - p + d (2 + (2 - 4 h) p) - d^2 (-1 + (1 - 2 h)^2 p) 
\end{matrix}\right]
\end{aligned}
\label{simple_z02}
\end{equation}

\begin{equation}
\begin{aligned}
y_{i} = 1: \mathbf{r_{i}^{2}} = \dfrac{1}{2} 
\left[\begin{matrix}
  1 - p + d (2 + (2 - 4 h) p) - d^2 (-1 + (1 - 2 h)^2 p),\\
\end{matrix}\right.\\
\qquad\qquad
\left.\begin{matrix}
   1 + p + d^2 (1 + (1 - 2 h)^2 p) + d (2 + (4h - 2) p) 
\end{matrix}\right]
\end{aligned}
\label{simple_z12}
\end{equation}

Similar to the 1-layer case, $\mathbf{(A+I)^{2} XW}$ can be represented as $\textbf{RW}$, where $\mathbf{R}$ is the matrix of updated features after two layers of feature aggregation. $\mathbf{RW}$ is then set equal to the label matrix $\mathbf{Y}$ after a transformation by the weight matrix $\textbf{W}$. Again, we sample the unique data points, leaving us with the optimal $\textbf{W}$ after solving the reduced system of equations.

We now consider a new test point, $t$ with label $y_{t} = 0$ and features $\mathbf{x_{t}} = \dfrac{1}{2} \begin{bmatrix} (1+p) & (1-p) \end{bmatrix}$. In addition, we assume that $t$ has a homophily ratio $h + \alpha_{t} = h_{t} \in [0, 1]$ where $\alpha_{t}$ represents a shift away from $h$. When $\alpha_{t}$ is positive, we can interpret this as a test node with a higher homophily ratio, and when $\alpha_{t}$ is negative, a lower homophily ratio. Under this new homophily ratio, $\mathbf{r_{t}^{2}}$ can be calculated as:

\begin{equation}
\begin{aligned}
\mathbf{r_{t}^{2}} = \dfrac{1}{2}
\left[\begin{matrix}
  1 + p + d^2 (1 + (1 - 2 h_{t})^2 p) + d (2 + (4h_{t} - 2) p),\\
\end{matrix}\right.\\
\qquad\qquad
\left.\begin{matrix}
   1 - p + d (2 + (2 - 4 h_{t}) p) - d^2 (-1 + (1 - 2 h_{t})^2 p) 
\end{matrix}\right]
\end{aligned}
\label{simple_zt2}
\end{equation}

We can now compute $\mathbf{r_{t}^{2}} \textbf{W} = \mathbf{z_{t}^{2}}$, the output of the 2-layer GNN, and analyze the associated predictions. The output of $\mathbf{r_{t}^{2}} \textbf{W}$ is shown in Equation \ref{equation_ztw2}, where $b_{2} = \dfrac{d}{(1 +d(2h-1))^{2}}$

\begin{comment}
\begin{equation}
\mathbf{z_{t}} = b_{2} \begin{bmatrix} 1 + 2 d (-1 + h + h_{t}) + d^2 (1 - 2 h + 2 h^2 - 2 h_{t} + 2 h_{t}^2) & 2 d (h - h_{t}) (1 + d (-1 + h + h_{t}))\end{bmatrix}
\label{equation_ztw2}
\end{equation}
\end{comment}

\begin{equation}
\begin{aligned}
\mathbf{z_{t}^{2}} = \dfrac{b_{2}}{d} 
\left[\begin{matrix}
  1 + 2 d (h + h_{t} - 1) + d^2 (1 - 2 h + 2 h^2 - 2 h_{t} + 2 h_{t}^2),\\
\end{matrix}\right.\\
\qquad\qquad
\left.\begin{matrix}
   2 d (h - h_{t}) (1 + d (-1 + h + h_{t})) 
\end{matrix}\right]
\end{aligned}
\label{equation_ztw2}
\end{equation}

We now consider what happens when $h_{t} = h + \alpha_{t}$ where $\alpha_{t} \in [-h, 1-h]$. Since we are interested in understanding how the predictions would change as a byproduct of $\alpha_{t}$, we look at the difference between predictions when $h_{t} = h$. Below we compute $\Delta \mathbf{z_{t}^{2}}$, the change in the logit vector output for the 2-layer GNN on the test point with local homophily $h_{t}$.

\begin{equation}
\begin{aligned}
\mathbf{z_{t}^{2}} = \dfrac{b_{2}}{d}
\left[\begin{matrix}
  1 + 2 d (h + (h + \alpha_{t}) - 1) + d^2 (1 - 2 h + 2 h^2 - 2 (h + \alpha_{t}) + 2 (h + \alpha_{t})^2),\\
\end{matrix}\right.\\
\qquad\qquad
\left.\begin{matrix}
   2 d (h - (h + \alpha_{t})) (1 + d (-1 + h + (h + \alpha_{t}))) 
\end{matrix}\right] 
\end{aligned}
\label{z_alpha}
\end{equation}

\begin{equation}
\begin{aligned}
&\mathbf{\Delta z_{t}^{2}} = 2b_{2} 
\begin{bmatrix}
  \alpha_{t} (1 + d (2 h - 1 + \alpha_{t})) & - \alpha_{t} (1 + d (2 h - 1 + \alpha_{t})) 
\end{bmatrix}  \\
&\mathbf{\Delta z_{t}^{2}} = 2(1 + d (2 h - 1 + \alpha_{t}))b_{2}
\begin{bmatrix}
  \alpha_{t} & - \alpha_{t} 
\end{bmatrix}  \quad \square
\end{aligned}
\label{z_alpha}
\end{equation}

The change in the logit vector, $\mathbf{\Delta z_{t}^{2}}$, for the 2-layer GNN, is similar to the change in the logit vector for the 1-layer GNN, except with an additional multiplicative factor. Specifically, if we refer to the change in the logit vector for the 1-layer GNN as $\mathbf{\Delta z_{t}^{1}}$, $\mathbf{\Delta z_{t}^{2}} = (2(1+d(2h - 1 +\alpha_{t}))\dfrac{b_{1}}{d}\mathbf{\Delta z_{t}^{1}}$. We focus on analyzing this multiplicative factor, given we already understand the behavior of $\mathbf{\Delta z_{t}^{1}}$ as studied in Section \ref{theory_section}. Our analysis is facilitated by letting the element-wise division of $\mathbf{\Delta z_{t}^{2}}$ and $\mathbf{\Delta z_{t}^{1}}$ be equal to $\mathbf{q} = \dfrac{\mathbf{\Delta z_{t}^{2}}}{\mathbf{\Delta z_{t}^{1}}} = \dfrac{2b_{1}(1+d(2h - 1 +\alpha_{t}))}{d} \begin{bmatrix}
  1 & 1
\end{bmatrix} = \dfrac{2(1+d(2h - 1 +\alpha_{t}))}{1+d(2h-1)} \begin{bmatrix}
  1 & 1
\end{bmatrix}
= \dfrac{2(1 + d(2h - 1) + d\alpha_{t})}{1+d(2h-1)} \begin{bmatrix}
  1 & 1
\end{bmatrix} $
and studying the coefficient of $\mathbf{q}$. Specifically, we analyze two extreme cases where $t$ is either part of a strongly heterophilous or strongly homophilous graph to see what patterns emerge. 

%TODO: Discuss solution, note that it follows similar to 1-layer case, but with added coeffecient. 

\noindent \textbf{Setting 1: Strongly Heterophilous $(\mathbf{\textit{h} \approx 0}$)}: We study this scenario by analyzing the limit of $\textbf{q}$ as $h$ approaches 0, with $d > 1$. 

\begin{equation}
    \lim_{h \to 0} \textbf{q} =  \dfrac{2(1 + d(\alpha_{t} - 1))}{1-d}
\end{equation}

In this scenario $0 \le a_{t} < 1$. When $d(\alpha_{t} - 1) < -1$, we see that the coefficient is positive, thus following the same degradation trend as in the 1-layer GNN. Conversely, when $d(\alpha_{t} - 1) > -1$ the coefficient is negative, indicating that the performance degradation does not occur. However, this requires that $\alpha_{t} > \dfrac{d-1}{d}$ which quickly becomes unlikely to occur as $d$ grows, as even low values of $d$ require extremely large $\alpha_{t}$ values to satisfy the inequality, e.g. $d = 4$ requires $\alpha_{t} > \dfrac{3}{4}$. We additionally analyze the magnitude of this coefficient, finding that $\dfrac{2(1 + d(\alpha_{t} - 1))}{1-d} > 1$ when $\alpha_{t} < \dfrac{d-1}{2d}$, suggesting that the 2-layer GNN more rapidly degrades in performance as compared to the 1-layer GNN when for moderate values of $\alpha_{t}$, and behaves similarly to the 1-layer when $\alpha_{t}$ is large and $d$ is small.

\vspace{0.1cm}
\noindent \textbf{Setting 2: Strongly Heterophilous $(\mathbf{\textit{h} \approx 1}$)}: We study this scenario by analyzing the limit of $\textbf{q}$ as $h$ approaches 1, with $d > 1$. 

\begin{equation}
    \lim_{h \to 1} \textbf{q} =  \dfrac{2(1 + d(\alpha_{t} + 1))}{1+d}
\end{equation}

In this scenario $-1 < a_{t} \le 0$. As in the 1-layer case, the coefficient is strictly positive, thus following the same degradation trend as in the 1-layer GNN. Similar to the heterophilous case, we find that $\dfrac{2(1 + d(\alpha_{t} + 1))}{1+d} > 1$ when $\alpha_{t} > -\dfrac{d-1}{2d}$, mirroring the behavior of the 2-layer GNN applied to heterophilous graphs with increased degradation. 

\noindent \textbf{High-level Takeaway:} When moving from a 1-layer to 2-layer GNN, we find that the trend of performance degradation still holds, with performance degrading as the local homophily ratio of a node deviates from the global homophily used to train the GNN. Additionally, we show that the rate of performance degradation for local homophily ratios far from the global homophily ratio is generally increased when a 2-layer GNN is applied, relative to a 1-layer GNN.

\subsection{Evaluation through Edge Homophily}
\label{app:other_metrics}
\rebuttal{In this section we discuss other edge-based homophily metrics proposed in the literature. Additionally, we discuss how our results generalize to higher-order neighborhoods. }

\rebuttal{ \noindent \textbf{Other Edge-based Homophily Metrics: } In this work, we focus on edge homophily as a means of characterizing performance. Recent work has pointed out that the global edge homophily metric, provided in Equation \ref{eq:global_homophily_ratio}, can become susceptible to imbalance in multi-class settings \cite{lim2021large}. To remedy this issue, the authors propose a class homophily metric which accounts for the size of each class within the global homophily calculation. As our analysis focuses on local homophily analysis, the issues in regards to the global homophily metric do not directly apply when stratifying across local homophily ranges. Additionally, the class homophily metric has no direct application to local analysis, given it depends on graph-level properties. While it is possible that the new class homophily metric can change the global homophily characterization, we find that the heterophilous datasets in this work have similar metrics as computed by both the class homophily and traditional global edge homophily \cite{lim2021large}, suggesting either metric would lead to similar conclusions. On heterophilous datasets with larger class imbalances, it may be prudent to use the class homophily metric to better contextualize discrepancy. }

\rebuttal{ \noindent \textbf{Higher-Order Behavior: } Two natural question arise as to whether the results hold when the GNN operates on higher-order information, as well as if the results can be contextualized by higher-order information. We first note that through the theoretical analysis in Appendix \ref{app:2_layer_proof}, we can see that a GNN operating on higher order information (2-hop neighborhoods) will experience similar degradation patterns. Given the relationship under the 2-hop setting is more complex, we prioritize the 1-hop theorem in the main text to provide a more intuitive understanding of how performance degrades. Moreover, all of the GNNs trained are cross validated on depths between 2 and 4, indicating that empirically the higher order information again clearly trends with our homophily measure. From the perspective of characterization, operating on higher-order neighborhoods often follows similar patterns as the one-hop case \cite{lim2021large, cavallo20222}, up to a certain point, as these metrics will approximate the global homophily when the size of the neighborhoods approach the entire graph. Moreover, there are practical challenges such as extremely large computation times, and ambiguity in metric design, that make it difficult to apply higher-order metrics. For instance, it is unclear if the metric should compare (a) the ego-node label with the labels for nodes in the $k$-hop neighborhood, (b) the labels between neighbors within the $k$-hop neighborhood, or (c) the ego-node label as compared with the labels of nodes exactly $k$-hops from the ego-node. In this work, we focus on the 1-hop case given its ubiquity across research on homophily/heterophily, as well as its intuitive definition. }

\subsection{Synthetic Datasets}

\subsubsection{Graph Generation Related Work}
\label{app:synth_dataset_related}
Karimi et al. proposed the homophily-based preferential attachment model that is used as the foundation for our synthetic generation method \cite{karimi2018homophily}. More recently, GraphWorld has been proposed to allow for the study of GNN performance across diverse graphs with varying global homophily, size, and class imbalance \cite{graphworld}. Additionally, FastSGG has been proposed to generate graphs for social network settings with an emphasis on control of the graph’s degree distribution \cite{fastsgg}. Despite these varying approaches, GraphWorld, FastSGG, and Karimi et al's preferential attachment model only consider homophily through a fixed global parameter, providing no insight into, or control of, the local homophily distribution in each graph. GenCAT remedies this issue by introducing a node-class matrix to control the homophily level for each node \cite{gencat}. However, while this matrix provides the flexibility to generate a graph with an arbitrary local homophily distribution, in practice there are no mechanisms proposed to enforce a particular distribution over it. Instead, each node's homophily is assumed to be completely independent, which is both unrealistic and cumbersome to instantiate. As such, systematically controlling this distribution would require the introduction of a parameterization over the node-class matrix, which is not proposed in the work. Our approach solves this problem by including a new \textit{single} parameter into the preferential attachment model that enforces nodes to deviate from the global homophily ratio, introducing nodes with a local homophily that has been shifted relative to the global homophily ratio. This capability allows us to easily learn over graphs with a wide range of local homophily ratios in a highly controlled manner, demonstrating the practical implications of our theory across different GNNs.

\subsubsection{Synthetic Generation Algorithm - Additional Details}
\label{app:synth_dataset_details}

\noindent \textbf{Initializing the Generated Graph:} We follow the same strategy to initialize the graph as is done in previous homophily-based preferential attachment models, as well as in the classic Barabasi-Albert (BA) algorithm \cite{Baraba_si_1999, karimi2018homophily, zhu2020beyond}. We begin by generating a node with label assigned from $P({0, ..., c})$ and add it to the empty graph $G$. We then sequentially add $m-1$ nodes to $G$, where each node is first assigned a class label and then attached to one other node in the graph. The attachment process for these nodes follows the GetNeighbors function as outlined in the main text, with only one edge added, rather than $m$ edges. The result of this process is a connected graph of $m$ nodes and $m$ edges, allowing any subsequent nodes to have $m$ connections without needing to create multiple edges between the same nodes. 

\begin{algorithm}[t!]
 \footnotesize
 \KwIn{Total nodes $n$, $m$ edges to add per step, label probability distribution $P(\{0, ..., c\})$, uniformity probability $\rho$, class compatibility matrix $\mathbf{H}_{L}$, drift change cutoff $\delta$} 
 Initialize $G$ with $m$ nodes and $m$ edges according to $\mathbf{H}_{L}$  \tcp*{Details in App. \ref{app:synth_dataset_details}} 
 Initialize vector $drift$ to hold change in homophily per node 
 $drift_{0}, ..., drift_{m} = 0$ \\
 \For{$u = m$ to $n$}{ 
        Add node $u$ to $G$ \\ 
      Sample label $y_{u} \sim P(\{0, ..., c\})$   \\
      $N_{1}(u)$ = GetNeighbors($G$, $m$, $u$, $drift$, $\rho$, $\mathbf{H}_{L}$, $\delta$) \\
       \For{$v$ in $N_{1}(u)$} {
            Add edge $(u, v)$ to $G$ \\
            \eIf{$y_{u}$ == $y_{v}$} {
              $drift[v]$ += 1 \tcp*{Drifted more homophilous}
              } {
             $drift[v]$ -= 1 \tcp*{Drifted more heterophilous}
              }
          }
      $drift[u] = 0$ 
 }
 \KwResult{Graph $G$ } 
 \caption{Synthetic Graph Generation}
 \label{synth_gen}
\end{algorithm}
 % \vspace{-0.4cm}

 % \vspace{-.3cm}
\begin{algorithm}[t]
\footnotesize
 \KwIn{$G$, $m$ edges to add, node $u$, $drift$ vector, uniformity probability $\rho$,  class compatibility matrix $\mathbf{H}_{L}$, drift change cutoff $\delta$}

\eIf(\tcp*[f]{Connect based on \textit{random} local homophily $h'$}){$q \sim U(0, 1) \le \rho$}  
{
  Sample new local homophily ratio $h' \sim U(0, 1)$ \\
  $hom\_drift = \{q : drift_{q} > \delta\}$  \tcp*{Get nodes with large homophilous drift}
  $het\_drift = \{s : drift_{s} < -\delta\}$  \tcp*{Get nodes with large heterophilous drift}
  $N_{1}(u)$ = \{Sample up to $round(mh')$ nodes with label $ y_{u}$ from $het\_drift$\} \\
  \If {$|N_{1}(u)|$ < $round(mh')$}{
    $N_{1}(u) \cup $ \{Sample $round(mh')$ - $|(N_{1}(u))|$ nodes with label $y_{u}$ from $G \setminus het\_drift$\} \\} 
  $N_{1}(u) \cup$ \{Sample up to $round(m(1-h'))$ nodes with label $ \ne y_{u}$ from $hom\_drift$\} \\
  \If {$|N_{1}(u)|$ < $m$}{
    $N_{1}(u) \cup$ \{Sample $m- |N_{1}(u)|$ nodes with label $ \ne y_{u}$ from $G \setminus hom\_drift$ \} \\}
}(\tcp*[f]{Connect based on original compatibility matrix $\mathbf{H}_{L}$}){
  %Initialize vector $scores$ = [] \\
 %\For{$v$ in $G$}{
  %    $scores[v]$ =  [$\mathbf{H}_{L}]_{y_u, y_v}$ \\}
  $N_{1}(u)$ = \{Sample $m$ nodes with probability based on $\mathbf{H_{L}}$\}
}
\KwResult{ $N_{1}(u)$ }
 \caption{GetNeighbors}
 \label{synth_neighbors}
\end{algorithm}

\subsubsection{Synthetic Dataset Properties}
\label{app:synth_dataset_props}
The synthetic dataset generation process is described in Section \ref{synth_generation_section} of the main text. Here, we detail some properties of the synthetic data, including the distribution of local homophily ratios as the uniformity parameter is varied and degree information.

\begin{figure}
    \centering
    \includegraphics[width=0.99\textwidth, keepaspectratio]{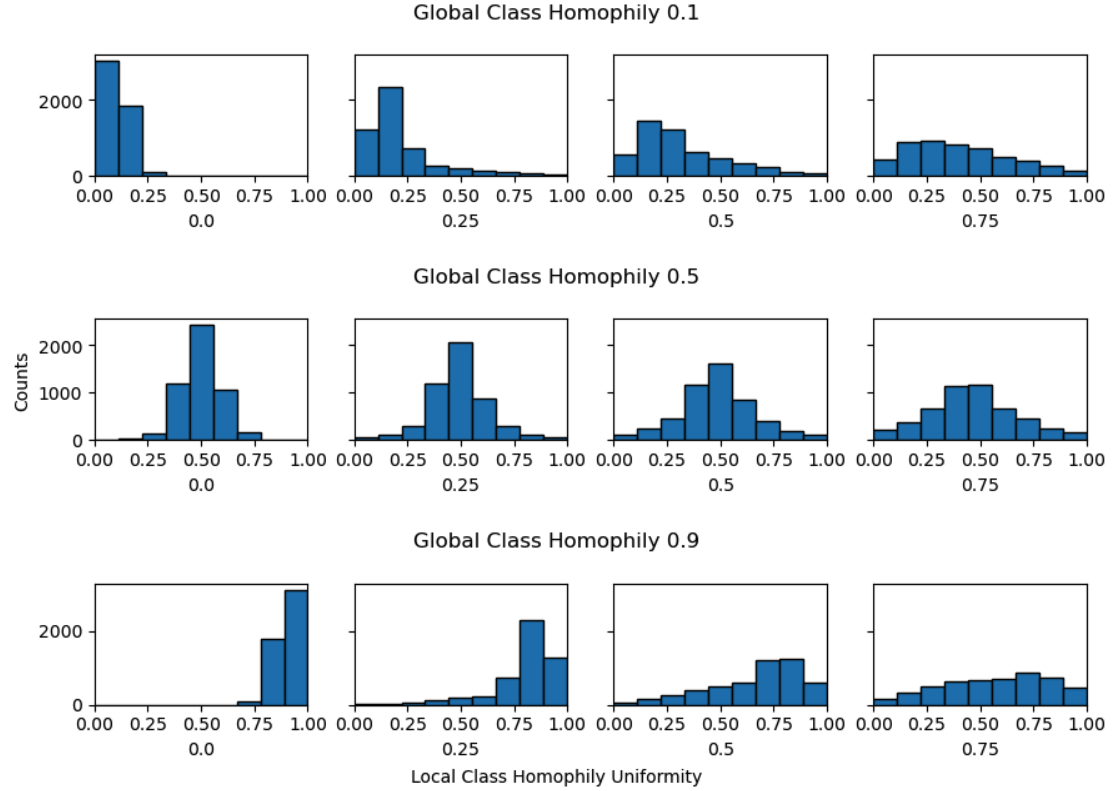}
    \caption{Distribution of local homophily ratios generated as the global class homophily is varied (rows) and the uniformity $\rho$ is varied (columns). We see the nodes spread out to occupy the full range of homophily values around $\rho = 0.5$ while still maintaining a clear peak.}
    \label{dist}
\end{figure}

\begin{figure}
    \centering
    \includegraphics[width=0.99\textwidth, keepaspectratio]{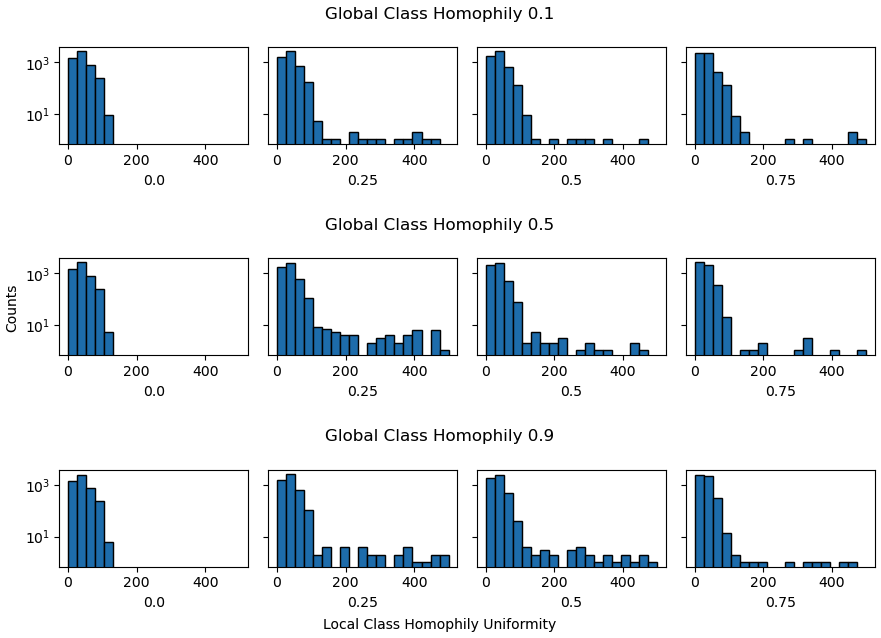}
    \caption{Distribution of degrees generated as the global class homophily is varied (rows) and the uniformity $\rho$ is varied (columns). We see that $\rho$ does not significantly impact the degree distribution other than for a small percentage of nodes.}
    \label{deg}
\end{figure}

In Figure \ref{dist}, we show how the distribution of local homophily changes as a function of the uniformity parameter. On the left-most plot for each row, we show the distribution of local homophily ratios on the unmodified preferential attachment model. Despite the minor deviation from the set global homophily ratio, the values are highly concentrated and do not provide samples across the entire range of local homophily values. This is similarly seen for the plot second to the left, which denotes $\rho = 0.25$, where the extreme homophily and heterophily values do not have enough nodes for analysis. The plot second from the right, $\rho = 0.5$, demonstrates the first scenario where there are ample data points across the the full range of local homophily and thus what we focus on in our analysis. Finally, the plot farthest to the right, $\rho = 0.75$, demonstrates a scenario where the distribution approaches uniformity. As we are interested in the problem of distribution shift at test time, this scenario is of less interest. 

In Figure \ref{deg}, we show the degree distribution for each of the graphs of varying uniformity level. As compared to the unmodified preferential attachment model, the uniformity parameter introduces a small amount of high degree nodes seen to the right of the distribution in each plot. This is due to the correction phase of the generation process, as described in the main text. However, as the uniformity parameter increases, the amount of high degree nodes becomes less severe. Moreover, even in the case of $h = 0.5, \rho = 0.25$, the high degree nodes only account for around $1\%$ of the nodes with the other graphs only containing even less.

\subsubsection{Additional Synthetic Results}
\label{app:synth_extra}

In this section, we include a series of additional experiments to supplement the findings in the main text. For our synthetic data, we include results for $h \in \{0.3, 0.7\}$, as well as additional models across all synthetic graph setups. 

\begin{figure}[h]
    \centering
  
    \includegraphics[ width=\textwidth, keepaspectratio]{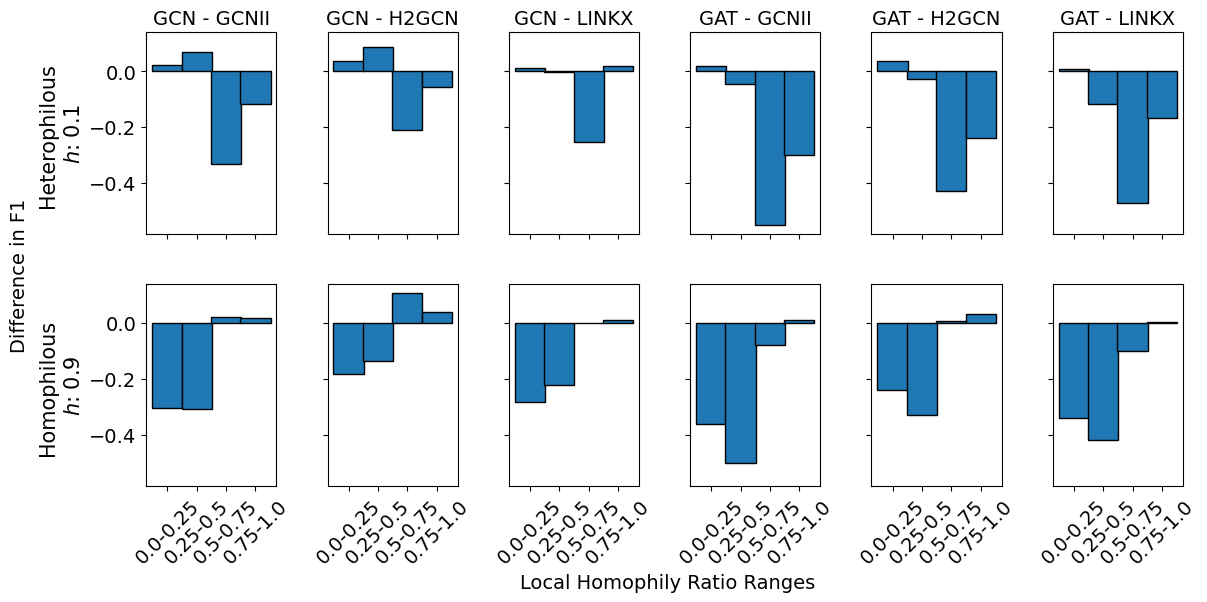}
    % \vspace{-0.35cm}
    \caption{Performance differences between homophilous and heterophilous models (columns) on synthetic data generated with global homophily ratios (columns) $h \in \{0.1, 0.9\}$ and uniformity $\rho = 0.5$. Each bar represents the average difference in F1 score between the two models specified in the column header for nodes with a local homophily ratio between the values specified on the x-axis.}
    \label{fig:pairwise_diff_synth}

\end{figure}

\noindent \textbf{Pair-wise Differences between Homophilous and Heterophilous Models.} In this section we re-plot Figure \ref{fig:local_vs_global_homophily}, directly subtracting the performance of the homophilous models from the heterophilous models. We show the difference in performance for all comparisons in Figure \ref{fig:pairwise_diff_synth}, where the homophilous model is subtracted from the heterophilous model. Thus, positive values are indicative of the homophilous model performing better, while negative values are indicative of the heterophilous model performing better. We show this for our two extreme settings, $h = 0.1, 0.9$. These plots visually reinforce our findings as the local homophily ranges near the global homophily ratio are always positive (homophilous models performing better), while the local homophily ranges far from the global homophily ratio are always negative (heterophilous models performing better). Additionally, the relative performance differences between the homophilous and heterophilous models in the positive and negative sections of Figure \ref{fig:pairwise_diff_synth} are significant -- heterophilous models achieve upwards of 0.45 F1 improvement in under-represented local homophily ranges while often losing less than 0.05 F1 on the over-represented local homophily ranges. 

\begin{figure}[H]
    \centering
  
    \includegraphics[ width=\textwidth, keepaspectratio]{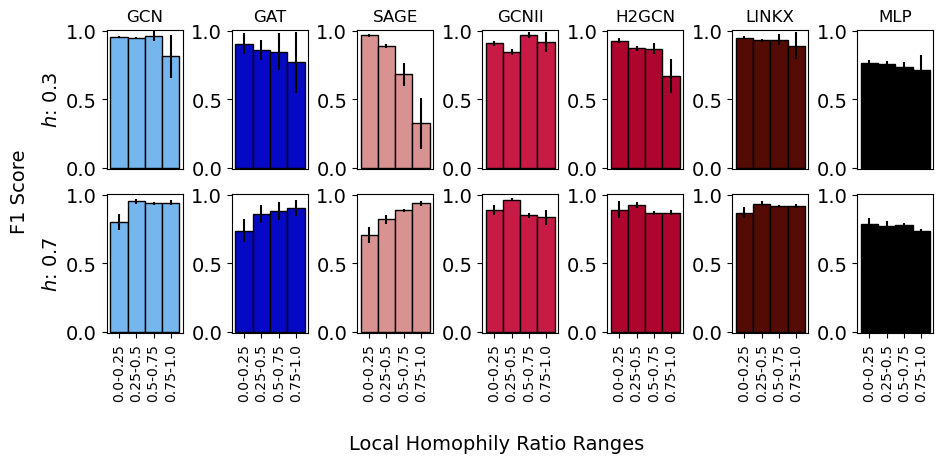}
    % \vspace{-0.35cm}
    \caption{Performance across models (columns) on synthetic data generated with global homophily ratios (rows) $h \in \{0.3, 0.7\}$ and uniformity $\rho = 0.5$. Each bar represents the F1 score for nodes with a local homophily ratio between the values specified on the x-axis. Error bars denote the standard deviation in performance. Blue bars indicate models with homophilous designs, red bar indicates models with heterophilous designs.}
    \label{fig:local_vs_global_homophily_37}

\end{figure}

\noindent \textbf{Additional Experiments for }$\mathbf{\textit{h} \in \{0.3, 0.7\}}$. In this section we highlight the performance of GCN, GAT, SAGE, GCNII, H2GCN, \rebuttal{LINKX}, and MLP models on synthetic settings where $h \in \{0.3, 0.7\}$. The results are shown in \ref{fig:local_vs_global_homophily_37}. While the discrepancy is not as significant as before, this is not unexpected as the theory indicates models will experience the most degradation at extreme local homophily shifts. However, almost every model, except for GCNII, degrades with a drop in F1 around 0.2 as the local homophily level shifts relative to the global homophily. Regarding model design, GCNII maintains a strong performance across all ranges in both settings, as does H2GCN in the $h = 0.7$ setting, showing the importance of heterophilous design to minimize disparity. 

\begin{figure}[H]
    \centering
  
    \includegraphics[ width=0.75\textwidth, keepaspectratio]{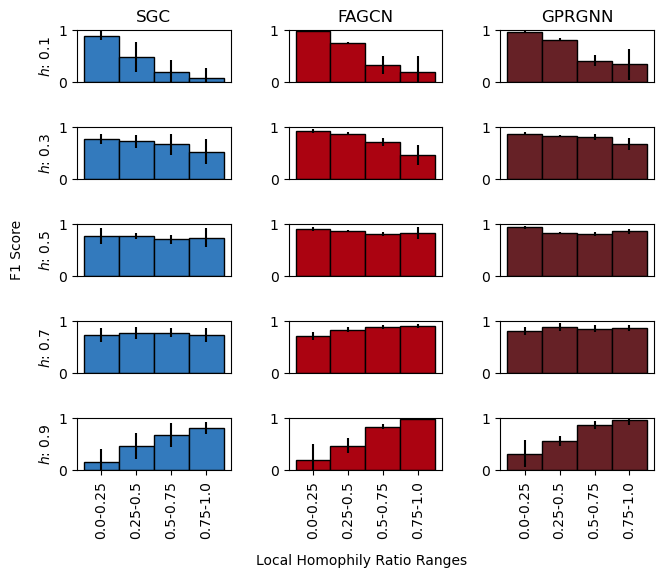}
    % \vspace{-0.35cm}
    \caption{Performance across models (rows) on synthetic data generated with global homophily ratios (columns) $h \in \{0.1, 0.3, 0.5, 0.7, 0.9\}$ and uniformity $\rho = 0.5$. Each bar represents the F1 score for nodes with a local homophily ratio between the values specified on the x-axis. Error bars denote the standard deviation in performance. Blue bars indicate models with homophilous designs, red bar indicates models with heterophilous designs.}
    \label{fig:synth_extras}
    \vspace{-.1cm}
    
\end{figure}

\noindent \textbf{Additional Models for All }$h$.
In addition to the models used in the main text, i.e. GCN, GAT, SAGE, GCNII, and H2GCN, we include results on SGC, FAGCN and GPRGNN. SGC is a linearized GNN architecture with a simple weighted aggregation tailored towards homophily settings. FAGCN and GPRGNN are both architectures designed to improve learning under heterophilous graphs, with the ability to differentiate homophilic and heterophilic nodes through gating mechanisms and adaptable weights, respectively. In Figure \ref{fig:synth_extras}, we show the results for these three models on each set of synthetic graph parameters. Similar to the main text, we find that SGC significantly degrades in performance as the local homophily ratio deviates from the global homophily ratio, even more so than GCN or GAT. Additionally, we see that FAGCN and GPRGNN both retain relatively strong performance, albeit slightly less than H2GCN and GCNII. We find that FAGCN and GPRGNN are much more sensitive to hyperparameters, as compared to H2GCN and GCNII, where the wrong choice can cause their performance to devolve close to that of GCN. While the general trend of heterophilous GNNs performing better on the full range of local homophily values remains true, H2GCN's concatenation based mechanisms provides significant practical value due to not relying on hyperparameters for strong performance.

\begin{figure}[H]
    \centering
  
    \includegraphics[ width=0.95\textwidth, keepaspectratio]{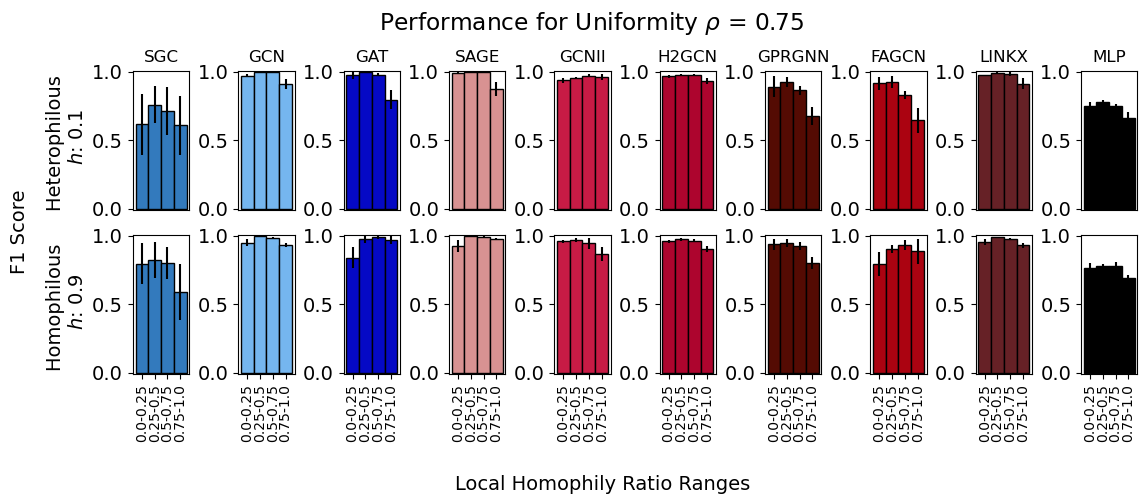}
    % \vspace{-0.35cm}
    \caption{Performance across models (columns) on synthetic data generated with global homophily ratios (rows) $h \in \{0.1, 0.9\}$ and uniformity $\rho = 0.75$. Each bar represents the F1 score for nodes with a local homophily ratio between the values specified on the x-axis. Error bars denote the standard deviation in performance. Blue bars indicate models with homophilous designs, red bar indicates models with heterophilous designs. By introducing more support across the range of local homophily ratios, each GNN is able to achieve much lower discrepancy across groups.}
    \label{fig:synth_diff_uniform}
    \vspace{-.1cm}
    
\end{figure}

\noindent \textbf{Additional Performance for $\rho = 0.75$}.
To further explain our choice for $\rho = 0.5$ in Section \ref{synth_section}, we train all of our models (GNNs and MLP) on additional synthetic datasets where $\rho = 0.75$. The distribution of local homophily ratios for $\rho = 0.75$ can be seen in Figure \ref{dist}. We do not consider $\rho = 0.25$ as there are not enough data points across the full local homophily range to achieve a reasonable signal, as seen in Figure \ref{dist}. For our experiments, we follow the same setup as in Section \ref{synth_section} by training each model on 10 different graphs of a particular parameter combination and analyze the performance trends across groups of nodes with varying local homophily level. We focus on $h = 0.1$ and $h = 0.9$ to maximize the possible shifts in local homophily, relative to $h$. The results are displayed in Figure \ref{fig:synth_diff_uniform} for all models. As expected, the performance discrepancy is significantly less than in Figure \ref{fig:local_vs_global_homophily}, across all model. Moreover, the difference in performance between the homophilous and heterophilous GNNs is less significant. As each GNN now has ample support across the full range local homophily range during training, each model is able to generalize, irrespective of the homophily level. While $\rho = 0.5$ is a useful tool to demonstrate how discrepancy can arise in GNN models, Figure \ref{fig:synth_diff_uniform} is valuable in showing a set of conditions that can allow GNNs to perform well, irrespective of design. 

\rebuttal{\noindent \textbf{Additional Discussion and Experiments on Training Set Size}. For our synthetic analysis, we use a training/validation/test split ratio of 50/25/25\%. Based on previous works for GNNs applied to \textit{heterophilous} settings, most use similar (or even higher) ratio to study their models \cite{zhu2020beyond, Pei2020Geom-GCN, hu2020open}. While studies have shown that GNNs applied to \textit{homophilous} settings can attain strong performance with low training node ratios \cite{kipf2016semi}, it is common for heterophilous graphs to require more training data as the local homophily patterns are more complex. Despite this known behavior, the impact of training size on disparity is not well studied. To address this gap, we include a set of experiments on our synthetic data where the amount of training data is varied. All of the parameters, aside from the amount of training data, is kept the same as described in Section \label{synth_section} of the main text. As the main text uses 50\% of the nodes for training, we study the scenarios of 30\%, and 10\%}. 

\begin{figure}[]
    \centering

    \includegraphics[ width=0.93\textwidth, keepaspectratio]{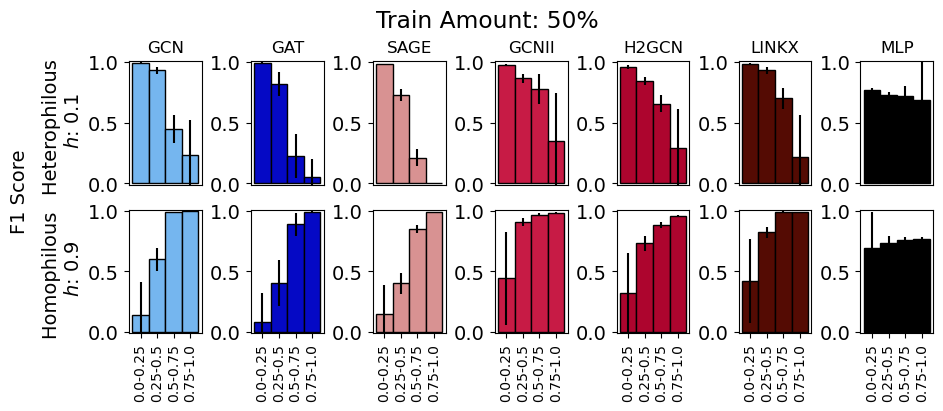}
  \includegraphics[ width=0.93\textwidth, keepaspectratio]{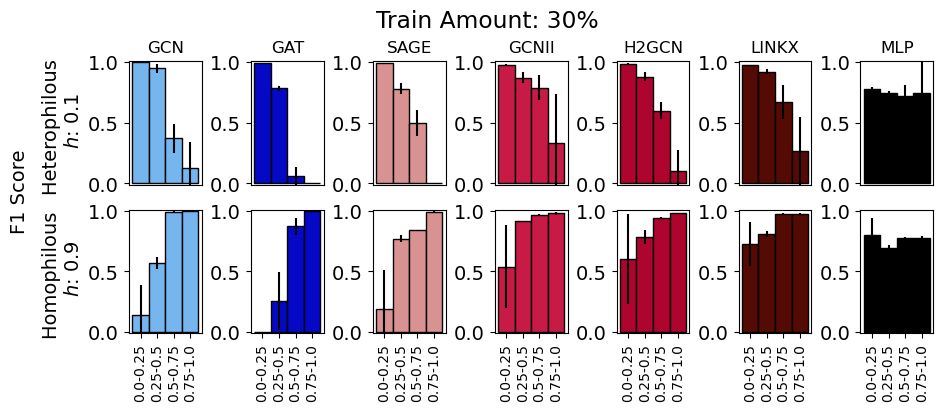}
    \includegraphics[ width=0.93\textwidth, keepaspectratio]{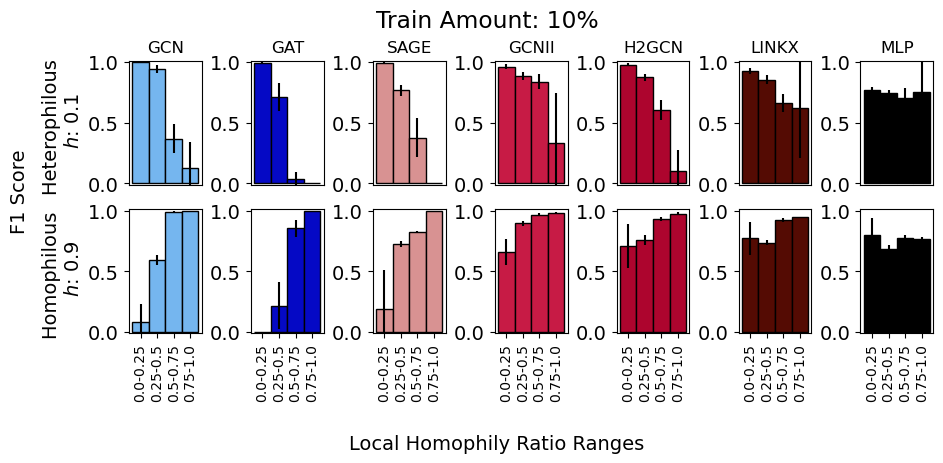}
    % \vspace{-0.35cm}
     \caption{\rebuttal{Performance across models (columns) on synthetic data generated with global homophily ratios (rows) $h \in \{0.1, 0.9\}$ and uniformity $\rho = 0.5$ with training set ratios of 50\% (top, the original results), 30\% (middle) and 10\% (bottom). Performance degradation patterns generally remain the same, highlighting discrepancy as a fundamental model and data distribution problem, rather than a direct artifact of training set size.}}
    \label{fig:train_amnt}

\end{figure}

\rebuttal{Through the experiments depicted in Figure \ref{fig:train_amnt}, we find that the amount of training data does not significantly alter the discrepancy patterns. This further verifies the findings of the main text that discrepancy is a fundamental issue embedded within the models and training distribution, rather than an artifact of training set size. We do note that many of the models have amplified degradation as the homophily level deviates from the global homophily ratio, as compared to the original results with a 50\% training set ratio. For example, GCN, GAT, and H2GCN nearly drop to an F1 score of 0 for the furthest bins from the global homophily ratio as the training set ratio drops. This is not unexpected, as the smaller training sets can cause less nodes to fall within these bins during training. Thus, training set size can \textit{influence} the distribution of local homophily levels, which impacts the discrepancy patterns, but it is not a direct issue. Some models retain comparable, or even slightly better, performance irrespective of training set size, however this is likely due to variations in training and are often accompanied by larger variance bars.}

\subsection{Real-world Datasets}

\subsubsection{Additional Models for Real-world Datasets}
\begin{figure}[h]
    \centering
  
    \includegraphics[ width=0.99\textwidth, keepaspectratio]{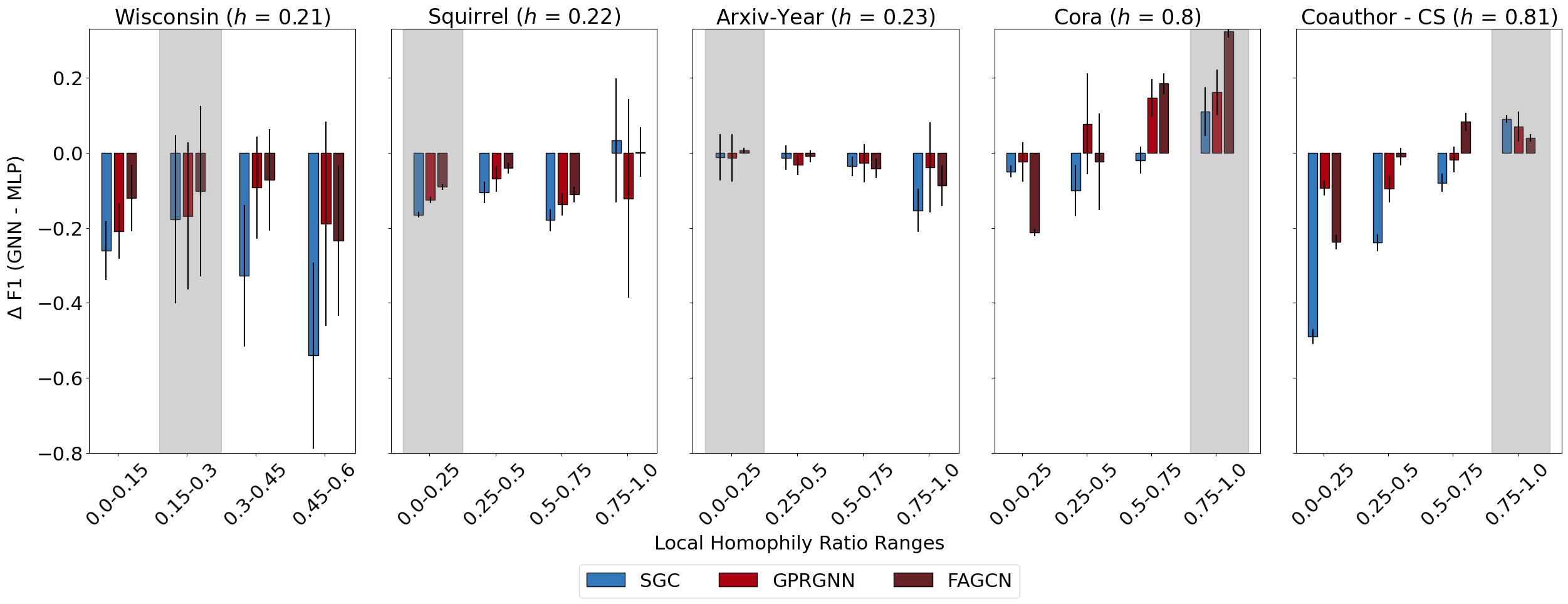}
    % \vspace{-0.35cm}
     \caption{F1 performance difference between GNNs and MLP ($\Delta$F1) for varying local homophily ratios in real-world graphs. For a homophily range specified on the x-axis, the 5 bars denote $\Delta$F1 for each GNN across test nodes that fall within the range (ranges adjusted per dataset such that each range has at least three nodes). Gray regions indicate the range that the global homophily ratio falls in. More negative bars indicate worse performance, while more positive bars indicate better performance. Error bars are the standard deviation in performance across experiment runs. Blue bars denote GNNs with homophilous designs, while red bars denote GNNs with heterophilous designs. All results align with previous findings where best performance is generally within the gray regions, and systematic performance disparity emerges as the ranges move further from the global homophily.}
    \label{fig:real_extras}
    \vspace{-.1cm}
    
\end{figure}

\label{app:real_world_extra}
Similar to our expanded analysis on the synthetic data, we include SGC, FAGCN, and GPRGNN results for the real-world datasets from the main text. As seen in Figure \ref{fig:real_extras}, the performance trends across the varying local homophily ranges generally stay consistent with the findings in the main text. In particular, we see that for Wisconsion, Arxiv-Year, Cora, and Coauthor-CS, the highest performing subgroups for each model is in, or near, the gray region, indicating that the global homophily ratio is influencing the performance. Squirrel demonstrates a slightly different behavior by not degrading at the furthest group, however does follow the trend across the first three groupings. In the main text, we note that there is noise in this regime due to the lack of nodes. Additionally, previous work has pointed out that the squirrel dataset can become easier to predict on due to duplicates across the structure \cite{platonov2023critical}. As such, it may be possible that the data leakage caused by these duplicates can influence performance, making certain homophily ranges easier.

\subsubsection{Local Homophily Distribution for Real-world Datasets}

For each the dataset analyzed in this work, we include a histogram of their local homophily distribution. While Cornell and Wisconsin are very similar and predominately heterophilous, the other datasets provide a wide spread of nodes across the varying local homophily ratios.  

\label{app:real_distr}
\begin{figure}[]
    \centering
  
    \includegraphics[ width=0.99\textwidth, keepaspectratio]{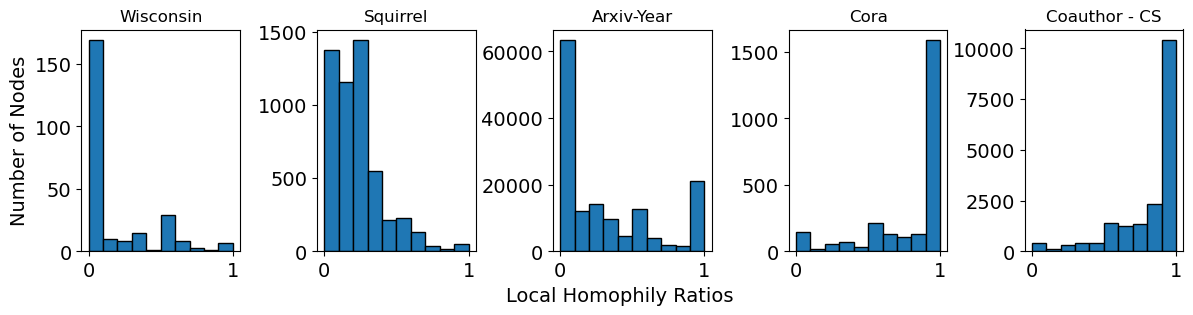}
    % \vspace{-0.35cm}
     \caption{Histograms depicting the distribution of local homophily ratios for the real-world datasets analyzed in the main text. Notably, all datasets, despite their clear peak, possess nodes across the entire range of possible local homophily ratios}
    \label{fig:real_properties}
    \vspace{-.1cm}
    
\end{figure}

\subsection{Models and Hyperparameter Tuning}
\label{app:hyp_param}
In this section we detail the models and hyperparameters tuned for each experiment. 

For synthetic experiments: 
\begin{enumerate}
    \item \textbf{MLP}: Implementation directly from PyTorch  
    \begin{itemize}
        \item Hidden Dim: 16, 32
        \item Depth: 2, 3
        \item Dropout: 0.3, 0.5
    \end{itemize}
    
    \item \textbf{GCN}: Implementation directly from PyTorch Geometric (torch\_geometric.nn.conv.gcn\_conv)
    \begin{itemize}
        \item Hidden Dim: 16, 32
        \item Depth: 2, 3
        \item Dropout: 0.3, 0.5
    \end{itemize}
    \item \textbf{GAT}: Implementation directly from PyTorch Geometric \\ (torch\_geometric.nn.conv.gat\_conv)
    \begin{itemize}
        \item Hidden Dim: 16, 32
        \item Depth: 2, 3
        \item Heads: 1, 2
        \item Dropout: 0.3, 0.5
    \end{itemize}
    \item \textbf{SAGE}: Implementation directly from PyTorch  Geometric (torch\_geometric.nn.conv.sage\_conv)
    \begin{itemize}
        \item Hidden Dim: 16, 32
        \item Depth: 2, 3
        \item Dropout: 0.3, 0.5
    \end{itemize}
    \item \textbf{GCN-II}: Implementation directly from PyTorch  Geometric (torch\_geometric.nn.conv.gcn2\_conv)
    \begin{itemize}
        \item Hidden Dim: 16, 32
        \item Depth: 4, 8
        \item Dropout: 0.3, 0.5
    \end{itemize}
    \item \textbf{H2GCN}: Open source PyTorch implementation from \\ https://github.com/CUAI/Non-Homophily-Large-Scale 
    \begin{itemize}
        \item Hidden Dim: 16, 32
        \item Depth: 2, 3
        \item Dropout: 0.3, 0.5
    \end{itemize}
    \item \textbf{SGC}: Implementation directly from PyTorch  Geometric (torch\_geometric/nn/conv/sg\_conv)
    \begin{itemize}
        \item Depth: 2, 3
    \end{itemize}
    \item \textbf{GPR-GNN}: Open source PyTorch implementation from \\ https://github.com/jianhao2016/GPRGNN
    \begin{itemize}
        \item Hidden Dim: 16, 32
        \item Depth: 2, 3
        \item Alpha: 0.1, 0.5, 0.9
        \item K: 10
        \item Dropout: 0.3, 0.5
    \end{itemize}
    \item \textbf{FA-GNN}: Open source PyTorch implementation from \\ https://github.com/bdy9527/FAGCN
    \begin{itemize}
        \item Hidden Dim: 16, 32
        \item Depth: 2, 3
        \item Epsilon: 0.1, 0.5, 0.9
        \item Dropout: 0.3, 0.5
    \end{itemize}
    \item \rebuttal{\textbf{LINKX}: Implementation directly from PyTorch  Geometric  \\ (torch\_geometric/nn/models/LINKX)}
    \begin{itemize}
        \item Hidden Dim: 16, 32
        \item Depth: 2, 3
        \item Dropout: 0.3, 0.5
    \end{itemize}
\end{enumerate}

For real-world experiments: 
\begin{enumerate}
    \item \textbf{MLP}: Implementation directly from PyTorch  
    \begin{itemize}
        \item Hidden Dim: 32, 64
        \item Depth: 2, 3, 4
        \item Dropout: 0.3, 0.5
    \end{itemize}
    \item \textbf{GCN}: Implementation directly from PyTorch Geometric (torch\_geometric.nn.conv.gcn\_conv)
    \begin{itemize}
        \item Hidden Dim: 32, 64
        \item Depth: 2, 3, 4
        \item Dropout: 0.3, 0.5
    \end{itemize}
    \item \textbf{GAT}: Implementation directly from PyTorch Geometric \\ (torch\_geometric.nn.conv.gat\_conv)
    \begin{itemize}
        \item Hidden Dim: 32, 64
        \item Depth: 2, 3, 4
        \item Heads: 1, 2
        \item Dropout: 0.3, 0.5
    \end{itemize}
    \item \textbf{SAGE}: Implementation directly from PyTorch  Geometric (torch\_geometric.nn.conv.sage\_conv)
    \begin{itemize}
        \item Hidden Dim: 32, 64
        \item Depth: 2, 3, 4
        \item Dropout: 0.3, 0.5
    \end{itemize}
    \item \textbf{GCN-II}: Implementation directly from PyTorch  Geometric (torch\_geometric.nn.conv.gcn2\_conv)
    \begin{itemize}
        \item Hidden Dim: 32, 64
        \item Depth: 4, 8, 16
        \item Dropout: 0.3, 0.5
    \end{itemize}
    \item \textbf{H2GCN}: Open source PyTorch implementation from \\ https://github.com/CUAI/Non-Homophily-Large-Scale 
    \begin{itemize}
        \item Hidden Dim: 32, 64
        \item Depth: 2, 3, 4
        \item Dropout: 0.3, 0.5
    \end{itemize}
    \item \textbf{SGC}: Implementation directly from PyTorch  Geometric (torch\_geometric/nn/conv/sg\_conv)
    \begin{itemize}
        \item Depth: 2, 3, 4
    \end{itemize}
    \item \textbf{GPR-GNN}: Open source PyTorch implementation from \\ https://github.com/jianhao2016/GPRGNN
    \begin{itemize}
        \item Hidden Dim: 32, 64
        \item Depth: 2, 3, 4
        \item Alpha: 0.1, 0.5, 0.9
        \item K: 10
        \item Dropout: 0.3, 0.5
    \end{itemize}
    \item \textbf{FA-GNN}: Open source PyTorch implementation from \\ https://github.com/bdy9527/FAGCN
    \begin{itemize}
        \item Hidden Dim: 32, 64
        \item Depth: 2, 3, 4
        \item Epsilon: 0.1, 0.5, 0.9
        \item Dropout: 0.3, 0.5
    \end{itemize}
    \item \rebuttal{\textbf{LINKX}: Implementation directly from PyTorch  Geometric  \\ (torch\_geometric/nn/models/LINKX)}
    \begin{itemize}
        \item Hidden Dim: 32, 64
        \item Depth: 2, 3, 4
        \item Dropout: 0.3, 0.5
    \end{itemize}
\end{enumerate}

All models have a final linear layer at the end of the convolution section to produce the final predictions. Any unspecified parameters, such as the additional parameters introduced by GCN-II, H2GCN, and LINKX are left as the defaults in their respective code bases.

\end{document}